# AI as a Democratizing Force in Indie Game Development:
## A Platform-Level Analysis Amid Industry Contraction, and the Coming Supply Surplus

**Brian Dean Madanamootoo Jatin Alla**

- *University of Silicon Valley, Gamers Home, Co-founder (California, United States)*
- *Kairos AI, Co-founder, (California, United States)*

## ABSTRACT

The video game industry of 2024–2026 presents a paradox: the deepest sustained contraction in its modern history at the AAA level, and the largest expansion of independent output ever recorded. This paper examines the role of artificial intelligence in that divergence through four research questions, grounded in three levels of evidence: public marketplace data (Steam, itch.io, industry surveys, funding trackers), a title-level Steam catalog dataset cross-referenced with Steam's generative-AI disclosure records, and a fourteen-month operational log from Gamers Home, an agentic AI production platform co-founded by the first author.

On barriers to entry (RQ1), we measure professional-grade production planning, historically the province of a salaried producer role costing roughly $59 per hour [1], being generated in a mean of 5.1 minutes at a cost of $0.27–0.58 per plan. We operationalize “democratization” across seven dimensions and claim it for one: the cost of the coordination function, repriced by roughly four orders of magnitude. The regional dimension of the question is argued from cost arithmetic and explicitly labeled as argued, not measured.

On market acceptance (RQ2), we document the indie expansion in volume, revenue share, and unit sales; an eightfold rise in AI-disclosed releases within eighteen months; and a professional sentiment collapse recorded across three consecutive industry surveys, alongside a title-level comparison of player reception showing disclosed releases received at catalog-typical rates (median 85.9% positive) in a verified subsample.

On whether generative AI has begun a new wave (RQ3), we situate the current moment as a third democratization wave following distribution (2008–2012) and construction (2013–2020), report four pre-registered convergence tests between platform and industry data, all aligned, and present the paper's central forward claim: when the cost of making games approaches zero while player demand in mature markets contracts, the market shows the preconditions of structural oversupply, forcing the emergence of a new distribution paradigm as waves one and two each forced their own.

On quality (RQ4), the platform log measures the generation of plans, not the quality of shipped games; the question is therefore answered from marketplace outcome data, comparing review-based quality signals for AI-disclosed titles against non-disclosed titles in a verified subsample, with the definitional boundary stated plainly: disclosure records generative-AI use in production, not authorship of the game.

We quantify both sides of the supply-demand imbalance (releases doubled from 9,654 in 2020 to over 20,000 in 2025 while only ~300 titles grossed above $1 million [2], and gamer participation in the eight largest markets fell below pre-pandemic levels [3]) and close with implications for developers, educators, platform designers, and the discovery layer this surplus will break.

---

---

# 1 INTRODUCTION

In 2026, the two most reported stories about the video game industry directly contradict each other.

The first is contraction: an estimated 4,600 industry layoffs in 2026 alone with twenty-two studio closures [4], following a documented wave of roughly 8,500 jobs in 2022, 10,500 in 2023, a peak of 15,650 in 2024, and 9,200 in 2025, approximately 44,000 positions in four years, 61% of them in North America [3].

The second is expansion: global game content sales reached an all-time high of $195.6 billion in 2025, up 5.3% year over year [3], while 10,676 independent games released on Steam in 2025, a 25% increase over the prior year and nearly double the 2022 figure [5]. Independent titles reached 48% of Steam's full-game sales revenue by 2024, double their 2018 share [6].

The industry is simultaneously shrinking and growing, depending entirely on where one looks: record revenue and record layoffs, in the same twelve months.

Artificial intelligence sits uncomfortably at the center of both stories. In the contraction narrative, AI is cast as a driver of job displacement: the Game Developers Conference's 2026 State of the Game Industry survey of more than 2,300 professionals (±3% margin of error) found that 52% now believe generative AI is having a negative impact on the industry, up from 30% a year earlier and 18% the year before that, a near-tripling of professional pessimism in two years, while positive sentiment fell from 21% to 13% to 7% across the same three surveys [7, pp. 3, 22]. In the expansion narrative, AI is cast as an enabler: the same survey found 36% of professionals actively using AI tools in their work [7, p. 19], and industry analyses increasingly credit AI-assisted workflows with making solo and small-team development viable at professional quality.

This paper does not attempt to adjudicate whether AI causes layoffs. That question, we will argue, is presently unanswerable with available evidence: companies cite restructuring, budget cuts, and market conditions rather than automation (only 6% of layoff-affected respondents' companies cited AI implementation as a reason [7, p. 16]), and named cases point in both directions. Instead, we pivot to questions that are measurable.

## 1.1 Research questions

**RQ1 (Barriers to entry).** Has generative AI reduced the barriers to entry for independent game development, specifically in regions with historically low indie game development rates? Operationally: has the cost and skill barrier to a core production input measurably fallen (measured), and does that fall plausibly extend access in low-participation regions (argued from cost arithmetic; per-project geography is unavailable in the current dataset and this sub-claim is labeled ARGUED, NOT MEASURED throughout, per §4.1.4)?

**RQ2 (Market acceptance).** How have games created or assisted by generative AI been accepted in the market? Acceptance is examined on three fronts: commercial acceptance by players (release volume, revenue share, unit sales, and review-based reception of AI-disclosed titles versus non-disclosed titles), professional acceptance by the industry (adoption and sentiment series), and the structural question of whether the expanded supply is being absorbed at all (saturation).

**RQ3 (A new wave).** Has generative AI begun a new wave of developing, distributing, or adopting games? Operationally: does the current moment fit the historical signature of a democratization wave (an output inflection at marketplace level, a cost collapse at tooling level, and a forced reorganization of distribution), and do platform-level and industry-level measurements converge on that reading?

**RQ4 (Quality).** What is the quality of the games that have been created with generative AI? The platform log measures plan generation, not shipped-game quality; this question is therefore answered from marketplace outcome data (§4.4), with the definitional boundary of §4.4.1: the measurable population is titles disclosing generative-AI use in production, which is not equivalent to games authored by generative AI.

## 1.2 Evidence

We answer with three levels of evidence and one sustained vantage point. At industry scale, we analyze Steam release data (2022–2026), itch.io platform milestones, GDC survey series, layoff tracking, and private-funding data. At title scale, we analyze a public Steam catalog dataset (§3.3) cross-referenced with Steam's generative-AI disclosure records, providing per-title quality and acceptance signals. At platform scale, we analyze a fourteen-month operational log (April 2025–July 2026) from Gamers Home, an agentic AI production platform on which independent developers generate structured production plans from design briefs.

The vantage point is the first author's: fifteen years building games and game businesses across two hemispheres, from founding and exiting Panda & Wolf Holding in Mauritius and shipping Eco-Warriors, the first mobile game to receive UNESCO patronage in Africa, to teaching game design at the University of Silicon Valley and co-founding the platform studied here.

## 1.3 Contributions

This paper makes five contributions, stated with their boundaries.

**C1 (empirical).** The first longitudinal, quantitative, project-level dataset of agentic AI production tooling in independent game development: fourteen months of operational telemetry covering plan generation speed, structural complexity (epics, stories), skill and tool identification, and marginal cost (Tables 4 and 19). This contribution stands independent of any causal claim.

**C2 (measurement).** A quantified repricing of one specific production function: the pre-production planning deliverable, from a $2,400–4,800, one-to-two-week producer-labor baseline to a measured 5.1 minutes and $0.27–0.58 (Table 6). This is a direct measurement of a mechanism, not an inference.

**C3 (conceptual).** An operationalized definition of production democratization as a construct distinct from the distribution democratization of wave one and the construction democratization of wave two (§2.2), with explicit dimensional boundaries (§2.1, Table 2) and a capability taxonomy for the tooling class it describes (§2.3, Table 3).

**C4 (predictive).** A falsifiable market-structure claim: that near-zero production cost plus collapsed capital gatekeeping, meeting contracting core-market demand, moves the market toward structural oversupply and forces a distribution-paradigm reorganization (§4.3.4), with registered falsification criteria (§5.7).

**C5 (outcome analysis).** A title-level cross-reference of verified generative-AI disclosure groups with catalog quality and acceptance signals (review ratios, ownership tiers), comparing disclosed releases against non-disclosed titles, executed as a verified subsample with the full-census matched design registered as the upgrade (§4.2.2, §4.4.2). To our knowledge no comparable title-level comparison exists in the published literature.

What this paper does not contribute: a causal identification of AI's effect on aggregate release volume (§4.3.3 presents the confounded evidence and rival explanations), a demonstration that democratized production improves outcomes for participants (§4.2.4 argues the median outcome may worsen), a comparison of plan quality between AI-assisted and traditional planning (§6, registered future work), or any claim about games authored end-to-end by generative AI, a population that current disclosure data cannot isolate (§4.4.1).

### 1.4 A note on the first author's position

The first author is a co-founder of the platform studied. We treat this methodologically rather than defensively: insider access is the reason this dataset exists at all, as no outside researcher could obtain fourteen months of operational logs from a live production platform. The first author writes from three vantage points that shape the analysis: platform co-founder, game-design educator, and developer from the Global South (Mauritius) who built under exactly the cost structures this paper measures the collapse of.

The controls on this conflict of interest are structural: registration of research questions, definitions, and methods prior to finalization of results [registration platform and DOI pending; see §3.5]; de-identification of all user data; the addition of title-level marketplace analyses (§3.3) whose data source is public and independent of the platform; and a commitment, honored throughout, to report findings unfavorable to the platform's own thesis (§4.1.5, §4.2.3, §4.4.3, §4.4.4).

## 2 BACKGROUND AND RELATED WORK

The claim that technology is "democratizing game development" is not new; it has been true, in a specific and bounded way, twice before. Each prior wave removed one barrier, left the next one standing, and forced the market to reorganize around the new constraint. Situating the current moment against those precedents is what separates structural analysis from hype, and it is what generates this paper's closing prediction.

**Table 1. The three democratization waves and their market consequences.**

| Wave | Period | Barrier removed | Enabling technology | Output signature | Reorganization forced |
|---|---|---|---|---|---|
| One | 2008–2012 | Publisher gatekeeping (distribution) | XBLA, PSN, Steam | Small-team titles enter mainstream charts (Braid, Super Meat Boy, Minecraft) | Digital storefronts replace retail shelf economics |
| Two | 2013–2020 | Tooling cost (means of building) | Unity free tier, Unreal rev-share, GameMaker, Godot, itch.io open publishing | itch.io passes 1,000,000 products by Nov 2024 [8]; Steam releases reach 9,654/yr by 2020 [2] | Steam Direct ($100 open door, 2017) replaces Greenlight curation |
| Three | 2023–present | Producer function (management of the build) | Generative AI (assets/code), agentic AI (planning, decomposition, ops); MCP infrastructure layer [9–11] | This paper's subject: §4.1 and §4.2 | This paper's prediction: §4.3.4 |

**Wave one (circa 2008–2012) democratized distribution.** Before digital storefronts, reaching players required a publisher, and publishers were gatekeepers of shelf space. Xbox Live Arcade, PlayStation Network, and above all Steam collapsed that gate. The releases that defined the first "indie boom," Braid (2008), Super Meat Boy (2010), Minecraft (2011), were made by teams of one to three people who could not have shipped a boxed product in 2005 [12]. The barrier removed was publisher gatekeeping. What it did not remove was the cost of making the game: engines were licensed at studio prices, and the tooling assumed professional teams.

**Wave two (circa 2013–2020) democratized the means of building.** Unity's free tier, Unreal's shift to revenue-share pricing, GameMaker, Godot, and the asset-store economy put professional-grade construction tools in the hands of anyone with a laptop. itch.io, launched in 2013 on an open-publishing model, became the wave's purest expression: over one million hosted products by November 2024 [8], including more than 557,000 entries from community game jams by mid-2026 [13]. On Steam, Valve's own reorganization for this wave, the 2017 replacement of curated Greenlight with the $100 open door of Steam Direct, took annual releases from a curated trickle to 9,654 by 2020 [2]. The barrier removed was tooling cost. What it did not remove was

operational overhead: the knowledge and labor of planning, scoping, decomposing, and managing a production, the work of the producer role, which funded studios staff and solo developers historically either learn slowly, buy expensively, or skip. This gap is observable at both ends of the industry: in emerging-market studios, where structured production management is priced entirely out of reach relative to local revenue (§4.1.4), and in game design education, where production planning remains the competency students most consistently lack on arrival (§5.4).

**Wave three (2023–present) is democratizing the management of the build.** Generative AI addressed assets and code first; agentic AI systems now address production itself: planning, decomposition, scheduling, role identification. The technical substrate of this wave is documented and industry-wide: the Model Context Protocol, open-sourced by Anthropic in November 2024 as a standard for connecting AI systems to external tools and data, reached over 10,000 public server implementations and 97 million monthly SDK downloads by March 2026, and passed to neutral Linux Foundation governance in December 2025 [9–11]. Agentic production tooling for game development is one vertical of that general shift. The barrier being removed is the producer function and the operational overhead of running a production.

## 2.1 Operationalizing "democratization"

"Democratization" is used loosely across the literature and the trade press, and a claim that AI "democratizes game development" is unfalsifiable until the dimension is specified. We therefore decompose the term and state, for each dimension, whether this study measures it, and what we find.

**Table 2. Dimensions of democratization: what this study does and does not claim.**

| Dimension | Definition | Measured here? | Finding |
|---|---|---|---|
| D1 Cost access | The monetary price of a production input falls | YES (primary) | Production-planning deliverable repriced ~4 orders of magnitude (Table 6) |
| D2 Skill access | The expertise required to obtain the input falls | PARTIALLY | Artifact requires no production-management training to generate; the judgment to evaluate it remains scarce (§4.1.5, §4.4.3). Artifact-level skill democratization claimed only |
| D3 Output volume | More games are made and shipped | OBSERVED, NOT ATTRIBUTED | Steam indie releases nearly doubled 2022–2025 (Table 7); attribution to AI is confounded (§4.3.3) |
| D4 Studio formation | More teams and studios form | NOT MEASURED | GDC self-funding (35%) and educator reports are suggestive context, not measurement |
| D5 Geographic participation | Participation broadens across regions and income levels | ARGUED, NOT MEASURED | Economic argument in §4.1.4; per-project geography unavailable; measurement registered as future work |
| D6 Creative diversity | The range of games widens | NOT DEMONSTRATED | Our own data suggests possible narrowing (homogenization clusters, §4.4.4); flagged as the decisive open question |
| D7 Commercial success | More participants achieve viable outcomes | EVIDENCE AGAINST | ~0.5–1.5% viability rates in a doubled supply pool (Table 13); §4.2.4 |

The paper's precise claim is therefore: wave-three tooling demonstrably democratizes dimension D1, and partially D2, for the production-planning function specifically; D3 co-moves but is not causally attributed; D5 is argued but not measured; D6 and D7 are not demonstrated, and the paper's own evidence suggests they may move in the opposite direction. Any reading of this paper as claiming that "AI democratizes game development" in the undifferentiated sense is a misreading the authors wish to prevent here.

## 2.2 Production democratization as a construct: boundaries

Production democratization is distinct from the two adjacent constructs it is most easily confused with, and the distinction is functional, not rhetorical. Decompose game development into three input classes: content (the assets that go into the game: art, audio, narrative text, code as material), construction (the tooling that assembles content into a running artifact: engines, editors), and coordination (the managerial function that decides what gets built, in what order, by whom, with what: planning, decomposition, scoping, role identification, scheduling).

Wave two democratized construction: engines became free, and asset stores partially commoditized content. Generative AI's first application to games (2022–2024) extended content democratization: image, audio, code, and text generation, which is where 60% of Steam's AI disclosures sit [14] and where the prior qualitative literature focused [15–18]. **Production democratization, the subject of this paper, is the democratization of coordination**: the producer function, previously acquirable only through salaried expertise, prolonged self-teaching, or omission.

The boundaries of the construct, stated explicitly: it covers the generation of coordination artifacts (plans, decompositions, role maps) and their marginal cost. It does not cover execution of the plan (the work itself remains undone), judgment about the plan (evaluation, editing, rejection; §4.4.3 shows this remains scarce), or outcomes of the plan (shipping, revenue; unmeasured in the platform log, §5.7). A study finding that plans are cheap is not a study finding that games are easy, and this paper claims the former.

## 2.3 Defining the tooling class: a capability taxonomy

“AI producer” is a marketing term in the trade and an undefined one in the literature. We define the class by capability level, so the studied system can be located precisely and claims scoped to its level.

**Table 3. Capability levels of agentic production tooling.**

| Level | Capability | Human role | Example behaviors |
|---|---|---|---|
| L1 Planning agent | Generates coordination artifacts from a brief: epic/story decomposition, skill and tool identification | Initiates, evaluates, edits, executes | Brief in, structured plan out |
| L2 Workflow orchestrator | Maintains and updates the plan against project state; assigns, sequences, tracks | Approves, executes | Task routing, progress tracking, re-planning |
| L3 Autonomous producer | Executes coordination decisions with delegated authority: hires, schedules, reprioritizes without per-decision approval | Supervises | Not observed in this study; largely speculative as of 2026 |

The system studied here (Arielle, Gamers Home) operates at L1 with partial L2 features (plan persistence and collaborator-marketplace attachment). Every quantitative platform claim in this paper concerns L1 behavior: the generation of the coordination artifact. The taxonomy also bounds the wave-three claim: the wave that is measurable today is an L1 wave; whether L2–L3 systems reproduce or amplify the economics measured here is registered future work. The MCP infrastructure figures [9–11] describe the substrate on which L1–L2 systems are being built across industries, not evidence of L3 deployment.

Each prior wave produced a measurable signature, and each forced a reorganization of how games reach players (Table 1, final column). If wave three is real, its signature should be visible in two places: an inflection in output volume at the marketplace level, and a collapse in the time and money cost of production planning at the tooling level. Sections 4.1 and 4.2 test for exactly those two signatures; Section 4.3 asks whether they constitute a wave; §4.3.4 argues the forced reorganization is already beginning.

## 2.4 Related Work

Four literatures border this paper; none yet covers its ground.

**Industry-trend reporting** provides the macro series this paper draws on: the GDC State of the Game Industry surveys (2025, 2026) for adoption and sentiment [7, 19]; SteamDB for release-volume data [5]; Ball/Epyllion for revenue, funding, and demand-side analysis [3]; and layoff tracking for the contraction series [4, 20]. This literature is quantitative but aggregate: it measures the industry, not the project.

**Qualitative studies of AI in game development** provide the closest academic precedent. Panchanadikar and Freeman's CHI PLAY 2024 study [15], recipient of a Best Paper Honorable Mention, is the nearest prior work: a qualitative analysis of 3,091 online posts and comments from indie-developer communities on Reddit and Facebook, examining how indie developers perceive and envision generative AI's role in their creative practice. The same group's follow-ups extend the line into 2026, examining generative AI as a quasi-teammate in small-scale creative teamwork [16] and its dual productivity and social role in indie developers' work [17]; a 2025 qualitative research synthesis consolidates the broader literature [18]. This body of work is project-adjacent but not quantitative, and it predates agentic production tooling specifically: its subjects used generative AI for assets, code, and ideation, not for production management.

**Agentic-AI infrastructure literature** documents the technical shift described in §2 [9–11] but contains no study of its application to game production.

**Market-structure economics of cultural industries**, the oldest of the four, predicts the dynamics this paper measures. Rosen's economics of superstars [21] shows formally that when the cost of serving additional consumers approaches zero, cultural markets concentrate rewards on a small number of winners regardless of marginal quality differences: exactly the ~300-of-20,000 concentration observed in Table 13. Anderson's long-tail hypothesis [22] offered the optimistic counter-prediction, that frictionless distribution would shift value into the niche tail; two decades on, the Steam data of §4.2.4 is closer to Rosen than to Anderson, with roughly half of releases earning effectively nothing. Salganik, Dodds, and Watts demonstrated experimentally [23] that in cultural markets with abundant choice, social influence rather than intrinsic quality determines which products win, and that winners are largely unpredictable: the mechanism underlying the virality-native pattern of Table 8 and the prediction of §4.3.4.

This paper's contribution to that literature is to supply the production-side series these models take as given: what happens to supply when its cost, not only distribution's, approaches zero.

To these four, the quality question of RQ4 exposes a fifth gap: no published analysis compares marketplace reception of AI-disclosed releases against matched non-disclosed releases at title level. Disclosure-count analyses exist [14]; matched outcome comparisons do not. §4.4 specifies that analysis.

The gap, then, is specific: no longitudinal, quantitative, project-level dataset of AI-assisted independent game production exists in the published literature, and no matched title-level reception comparison of AI-disclosed releases exists either. Industry data is aggregate; academic data is qualitative; infrastructure data is technical. This paper contributes the missing layers, connects them to the other literatures, and uses the combination to state where the market goes next.

# 3 DATA AND METHODS

## 3.1 Industry-level sources

Steam release volume is drawn from SteamDB's release statistics filtered to the Indie tag, retrieved July 1, 2026 [5]. SteamDB is a primary aggregator of Steam's own catalog data and the standard source for release-volume analysis. Its limitation is definitional: the Indie tag is community- and developer-applied, so the series measures tagged releases, not a curated census of independent development.

Professional adoption and sentiment figures come from the GDC State of the Game Industry surveys for 2025 and 2026 (2,300+ respondents, ±3% margin of error [7, p. 3]). Layoff figures come from Ball/Epyllion's annual analysis [3], cross-checked against community trackers [4, 20], with the caveat that layoff tracking aggregates public reporting and therefore undercounts unannounced departures and overweights large studios. Revenue, funding, and demand-side figures come from Ball/Epyllion [3], VG Insights [6], and Alinea Analytics [2].

itch.io, the platform most associated with wave-two open publishing and the first author's primary distribution channel across a decade of indie practice, publishes no year-by-year release series and has stated it will not release per-project statistics [24]. We therefore use itch.io milestone figures (one million products by November 2024 [8]; 557,000+ jam entries by mid-2026 [13]) as context, and Steam as the sole longitudinal storefront series. This is a real limitation, discussed in §5.7: Steam's series likely under-represents the most informal tier of independent output, which is precisely the tier a democratization thesis concerns, meaning our supply-side numbers are conservative floors.

## 3.2 Platform-level source: the Gamers Home operational log

Gamers Home is a production platform for independent developers, co-founded by the first author, whose core system, an agentic AI producer ("Arielle," an L1 planning agent with partial L2 features per Table 3), converts a design brief or game design document into a structured production plan: a hierarchy of epics decomposed into stories, with required skills and tools identified per project, attached to a collaborator marketplace spanning 38 countries with 900 enrolled students via .edu access. The platform's architecture is representative of the agentic-tooling category described in §2.3 rather than unique to it, and is treated as a measurement instrument for the category: the measured quantities (generation time, plan scale, marginal cost) are properties of the underlying agentic-model architecture that any comparable L1 system inherits.

The dataset is an operational event log covering April 2025 through July 2026, approximately fourteen months, containing roughly 213 project-creation events across approximately 189 unique project titles. This log is a sample of a larger operational corpus [approximately 3x larger, referent under confirmation].

**Table 4. Dataset characteristics: the Gamers Home operational log. Source: [25]. All figures aggregate; no user-level data reported.**

| Characteristic | Value | Notes |
|---|---|---|
| Observation window | April 2025–July 2026 (~14 months) | Timestamps recovered from system identifiers |
| Project-creation events | ~213 | Unit of count for genre/category analysis |
| Unique project titles | ~189 | Events exceed titles due to re-creation and iteration |
| Internal/demo/test accounts | ~27% of creation events | Excluded from adoption claims; retained for system metrics |
| Fully-logged plan generations (timing) | n = 40 | Basis for speed statistics (Table 19) |
| Cost-logged plan generations | n = 64 (34 earlier, 30 later batch) | Basis for cost statistics (Table 19) |
| Mean plan size | 15.8 epics / 59.1 stories | Maximum observed: 33 epics / 122 stories |
| Project category split | ~2/3 game design, ~1/3 studio operations | §4.1.2 |
| Genre classification | Keyword-derived, multi-label, event-counted | Full-sample distribution in Table 5 |
| Sample relation to full corpus | ~1/3 of a larger operational corpus | Referent under confirmation |
| De-identification | Complete prior to analysis | Emails, names, conversational content excluded |

Five methodological decisions govern its use.

**De-identification.** All user-identifying information (email addresses, names, and conversational content present in the raw log) was excluded prior to analysis. Only aggregate statistics are reported. The de-identification protocol is described in the study's registration.

**Internal-account tagging.** Approximately 27% of creation events originated from platform-team, demonstration, or test accounts. These are flagged and excluded from any claim about external adoption. They are retained for system-performance metrics (plan-generation speed, output scale, and cost), which are properties of the system and valid regardless of who triggered generation.

**Genre methodology.** The log sample does not contain structured genre metadata. Genre classification was therefore derived by keyword matching against project titles, is multi-label (a title can match more than one category), and counts creation events rather than deduplicated games. The keyword lists are included verbatim in the study registration. This is the weakest measurement in the paper and is treated as such throughout: genre figures are reported as directional signal, and §5.7 specifies the upgrade path (developer-assigned tags on the full corpus).

**Scope.** All project types are retained, including non-game projects (marketing plans, business infrastructure, media production, internal tooling), reflecting the platform's observed use across the full lifecycle of an independent studio rather than only its game-design layer. This decision is itself a finding, reported in §4.1.2.

**Sampling.** Because the log is a sample of a larger corpus, absolute counts are lower bounds and proportions are estimates. No claim in this paper depends on the sample being exhaustive.

### 3.3 Title-level source: Steam catalog dataset with AI-disclosure cross-reference

To answer RQ2 (player-side acceptance) and RQ4 (quality) at title level, this study adds a third data source independent of the studied platform: the Steam Games Dataset published by FronkonGames [26], an openly licensed catalog scrape of Steam's public store and API records covering more than 100,000 titles, including per-title release date, price, review counts (positive/negative), Metacritic score where present, recommendation counts, average and median playtime, estimated ownership tiers, developer and publisher records, and genre classifications. The dataset version analyzed will be pinned by repository commit hash in the study registration for reproducibility (revision `dc0dcabe`, 2026-07-02, 134,712 titles, at this draft).

**Disclosure cross-reference.** The catalog dataset contains no generative-AI field. AI-disclosure status is therefore obtained by cross-referencing title AppIDs against Steam's generative-AI disclosure records, the same source underlying the disclosure-count series of Table 9 [14] [cross-reference source to be finalized: Lambe/Totally Human disclosure list, SteamDB AI-content tag export, or direct storefront-API sampling; the chosen source and its coverage limitations will be documented here].

**Analysis design (registered in advance of execution).** Population: Steam releases dated January 2024 (the start of Steam's disclosure requirement) through December 2025. Comparison: titles with a generative-AI disclosure versus non-disclosed titles matched on release year, price band, and genre. Matching prevents the comparison from reducing to composition effects (disclosed titles skewing toward low-price, high-volume genres). Quality and acceptance metrics, in registered order of primacy: (1) positive-review ratio among titles with at least 10 reviews (primary; near-universal coverage; the 10-review floor guards against small-denominator noise); (2) median playtime (behavioral engagement signal); (3) estimated-ownership tier (market-acceptance signal, bridging RQ2); (4) Metacritic score where present (secondary; sparse coverage, and a Metacritic value of 0 in the dataset denotes absence of a review, not a score, a zero-inflation property that applies to several fields and is handled by treating zeros in critic-score and playtime fields as missing where the field semantics require it).

Known data-quality constraints, stated in advance: the dataset's community-tag field is unpopulated in the analyzed distribution, so genre matching uses the populated genre field; review counts are cumulative at scrape time, so recently released titles carry fewer reviews (the matched design and review floor mitigate but do not eliminate this); and disclosure is self-reported and loosely enforced, making the disclosed group a floor, not a census [14].

**The definitional boundary, stated here and enforced in §4.4.1:** Steam's disclosure records generative-AI use anywhere in production, and 60% of disclosures concern visual asset generation [14]. The measurable population is therefore "titles disclosing generative-AI use," not "games created by generative AI." All RQ4 findings are scoped accordingly.

**Executed analysis and deviation from the registered design.** The comparison reported in Tables 10 and 17 was executed in a verified-subsample form that deviates from the full-census design above, and the deviation is documented here rather than silently absorbed. (1) *Disclosure source.* Full disclosure-record exports were unavailable at execution; disclosure groups were instead taken from the public replication package of an independent study of AI disclosure and player perception [27]: 424 post-2024 releases with store-page disclosure verified by that study's keyword protocol, 80 pre-2024 releases that added disclosure retroactively, and 93 non-disclosed controls, with no overlap between groups. The 424-title group was randomly subsampled to 200 (seed 42) for retrieval-budget reasons. (2) *Metrics source.* Per-title metrics were retrieved from a SteamSpy-derived catalog dataset [28] rather than [26], whose row payloads exceeded the retrieval channel available at execution; playtime and Metacritic fields are unpopulated in that distribution, so the registered metrics (2) and (4) are deferred and ownership tier substitutes as the behavioral signal. (3) *Coverage.* Match rates against the metrics dataset were 86/200 (disclosed), 71/80 (retro-labeled), and 54/93 (control); the dataset undercovers very recent and very small titles, biasing all groups toward visibility. (4) *Matching*. Fresh matching on release year, price band, and genre was not possible without the full catalog; group comparability is inherited from the source study's construction, with a price-band post-stratification check reported alongside the primary test. (5) *Selection*. Both disclosed and control groups were originally selected conditional on review volume, so the registered metric "share of titles reaching the 10-review floor" is not interpretable in this sample and is not reported. The full-census matched design remains registered as the upgrade path, with the pinned [26] revision above. Analysis code, group AppID lists, retrieved metric extracts, and the sampling seed are included in the deposit package.

### 3.4 Analytical approach

Industry-level analysis is descriptive trend analysis. Platform-level analysis is descriptive statistics on plan generation and classification of project content. Title-level analysis is the verified-subsample comparison executed per §3.3, with the full-census matched comparison registered as the upgrade path.

The connection between the platform and industry levels is made through **convergence analysis**: four pre-specified alignment tests (timing, genre, team-scale, and scope, defined in §4.3.1 and registered in advance) between independent datasets. We state plainly that this is not statistical correlation; a single platform's fourteen-month log against a five-year industry curve does not contain enough time-aligned observations for one, and pretending otherwise would be the kind of overreach this paper is designed to avoid. Convergence analysis asks a more modest question with an honest answer: do independent measurements at two scales point the same direction?

The forecast in §4.3.4 extrapolates the 2022–2026 Steam series through 2027 under two models, a linear trend and a structural-break model treating 2024 as a regime change, with both reported. Rival explanations for the industry series are treated systematically in §4.3.3. The supply-demand analysis of §4.3.4 combines the Steam supply series with Ball/Epyllion's demand-side participation data [3].

### 3.5 Ethics, registration, data availability, and conflict of interest

This study is a secondary analysis of pre-existing operational platform data and public marketplace data involving no intervention, interaction, or collection of data from human subjects for research purposes; under common institutional criteria (e.g., US 45 CFR 46.104(d)(4)) such analysis of de-identified records is exempt from IRB review. No author holds an institutional appointment requiring board submission for exempt secondary analyses; this determination is stated for transparency. Reporting is aggregate-only for platform data.

**Data availability.** The de-identified aggregate statistics underlying all platform tables, the genre keyword lists, the title-level analysis scripts, and the pinned identifier of the public catalog dataset version are deposited as supplementary materials with the study registration. Raw operational logs are not shared, as they contain user-identifying information that de-identification at the record level cannot fully remove. The Steam catalog dataset [26] is public.

**Conflict of interest.** The first author's conflict of interest is declared in §1.4 and treated as a methodological condition with structural controls, of which the registration is primary and the platform-independent title-level analysis (§3.3) is a second.

## 4 RESULTS

### 4.1 RQ1: Barriers to Entry

Has generative AI reduced the barriers to entry for independent game development, and does the reduction extend to regions with historically low participation? This section measures the mechanism directly (§4.1.1–§4.1.3), argues the regional extension from cost arithmetic under an explicit ARGUED, NOT MEASURED label (§4.1.4), and closes with the boundary of the claim (§4.1.5).

#### *4.1.1 The mechanism, measured*

The industry curve (§4.2.1) shows that independent output inflected; it cannot show why at the level of an individual creator's decision to start a project. The platform log can: it records, with timestamps and system telemetry, what happens when the production-planning barrier is removed. The analysis in this section draws on direct operational access to the platform, per the positionality and controls stated in §1.4 and §3.5.

Across forty fully-logged plan generations in the sample, the system produced a complete structured production plan in a mean of 308.7 seconds (median 310.4s; range 194–493s), approximately 5.1 minutes. The plans are not sketches: the mean plan comprises 15.8 epics decomposed into 59.1 stories, with the largest reaching 33 epics and 122 stories, alongside per-project identification of required skills and tools. In format and granularity, this is the artifact a staffed production team produces in pre-production, the epic and story decomposition of professionally managed agile development. The cost side is equally specific: mean $0.58 per plan across the earlier-logged batch (n=34) and $0.27 across the later batch (n=30), the marginal cost of professional-grade production planning roughly halving within the sample period as underlying model efficiency improved.

Two independent barrier-lowering mechanisms are therefore visible in the same log: the time cost collapsed by orders of magnitude relative to manual practice, and the dollar cost is both nominal and falling. The full instrumented sample is reported in Table 19 (Appendix A). Summary statistics: epics mean 15.8 (median 17, range 5–33); stories mean 59.1 (median 60, range 16–122); stories per epic mean 3.7; skills identified mean 3.2 (range 0–7, n=36); tools identified mean 4.2 (range 2–5, n=36); cost per plan mean $0.58 (n=34); cost per story mean $0.010, roughly one cent of planning cost per user story.

#### *4.1.2 What gets built*

Genre distribution across the full sample of creation events, under the registered keyword methodology of §3.2 (multi-label, creation events, directional), is reported in Table 5 and Figure 1.

**Table 5. Genre distribution, full platform sample (keyword-derived, multi-label, creation events). Source: [25]. Methodology per §3.2; directional signal only.**

| Category | Creation events |
|---|---|
| Simulation / Cozy | 20 |
| RPG / Fantasy | 20 |
| Roguelike / Roguelite | 17 |
| Action | 15 |
| Horror | 13 |
| Platformer | 12 |
| Strategy / 4X | 7 |

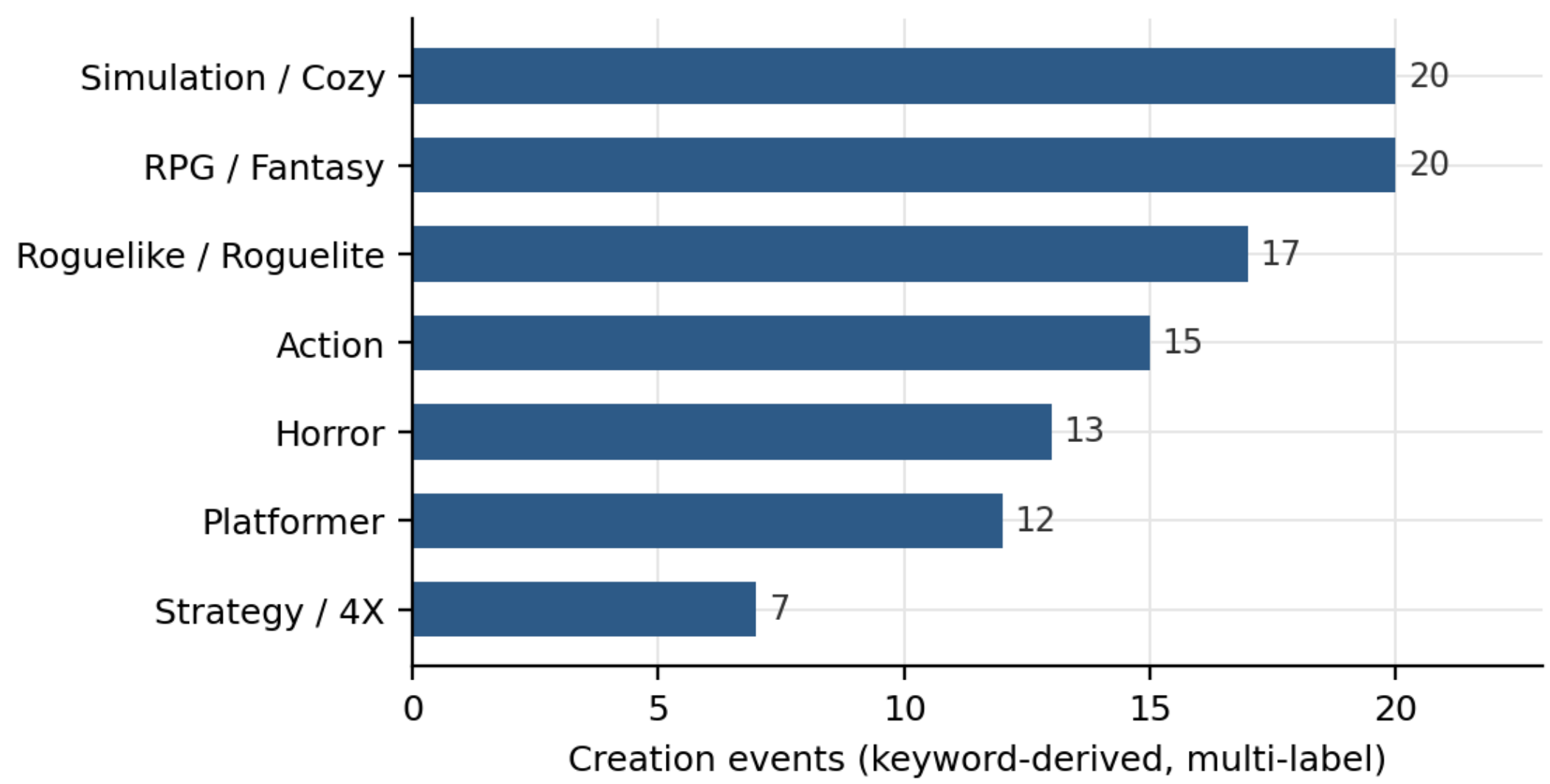

Figure 1. Genre distribution of platform creation events (keyword-derived, multi-label). The profile matches the contemporary indie genre cluster driving the Steam expansion of §4.2.1.

The profile matches the contemporary indie profile: cozy simulation, roguelikes, horror, and 2D platformers, the same genre cluster driving the Steam expansion of §4.2.1, rather than the AAA genre profile of open-world action and live-service shooters. The platform's pipeline previews the storefront. Within the instrumented subsample, plan complexity varies by game type (Table 20, Appendix A).

The second distributional finding was unanticipated at study design and is arguably more interesting: approximately **one-third of all in-scope creation events were not game-design projects at all** but studio operations: marketing and communications plans, media production (trailers, films, music), business infrastructure (financial planning, IT, compliance readiness), and internal tooling (Figure 2). Independent developers, given an agentic producer, use it the way a funded studio uses its operations staff, across the whole business, not only the game. This is the platform-scale reproduction of the GDC asymmetry noted in §4.2.3 (AI adoption highest in business roles, 58%), and it recurs as a load-bearing datum in the synthesis.

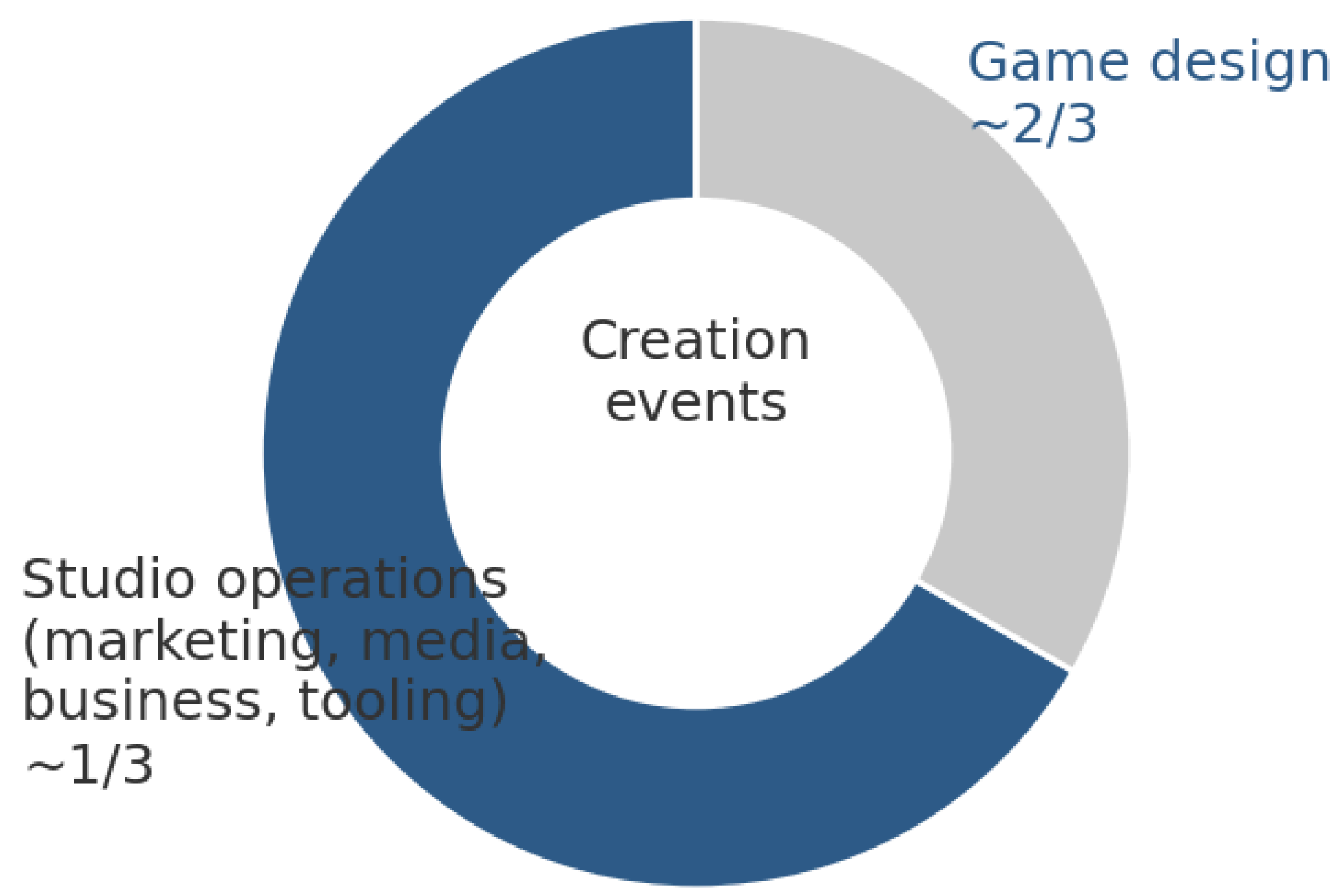

Figure 2. Project category split across in-scope creation events: approximately two-thirds game design, one-third studio operations.

#### *4.1.3 The barrier, repriced*

**Table 6. The cost of a production plan: traditional versus agentic.**

| Input | Traditional (US baseline) | Agentic (measured, GH log) | Ratio |
|---|---|---|---|
| Labor rate | $59.40/hr (producer avg $123,552/yr) [1] | n/a | |
| Time | 1–2 weeks of producer time (conservative for a 16-epic/59-story plan) | 5.1 minutes mean | ~700–1,400x |
| Direct cost | $2,400–4,800 before overhead | $0.27–0.58 | ~4,000–18,000x |
| Availability | Requires a hire; 50% of indie-studio staff earn under $100K vs 85% of AAA staff over it [29] | Any account, any geography | |

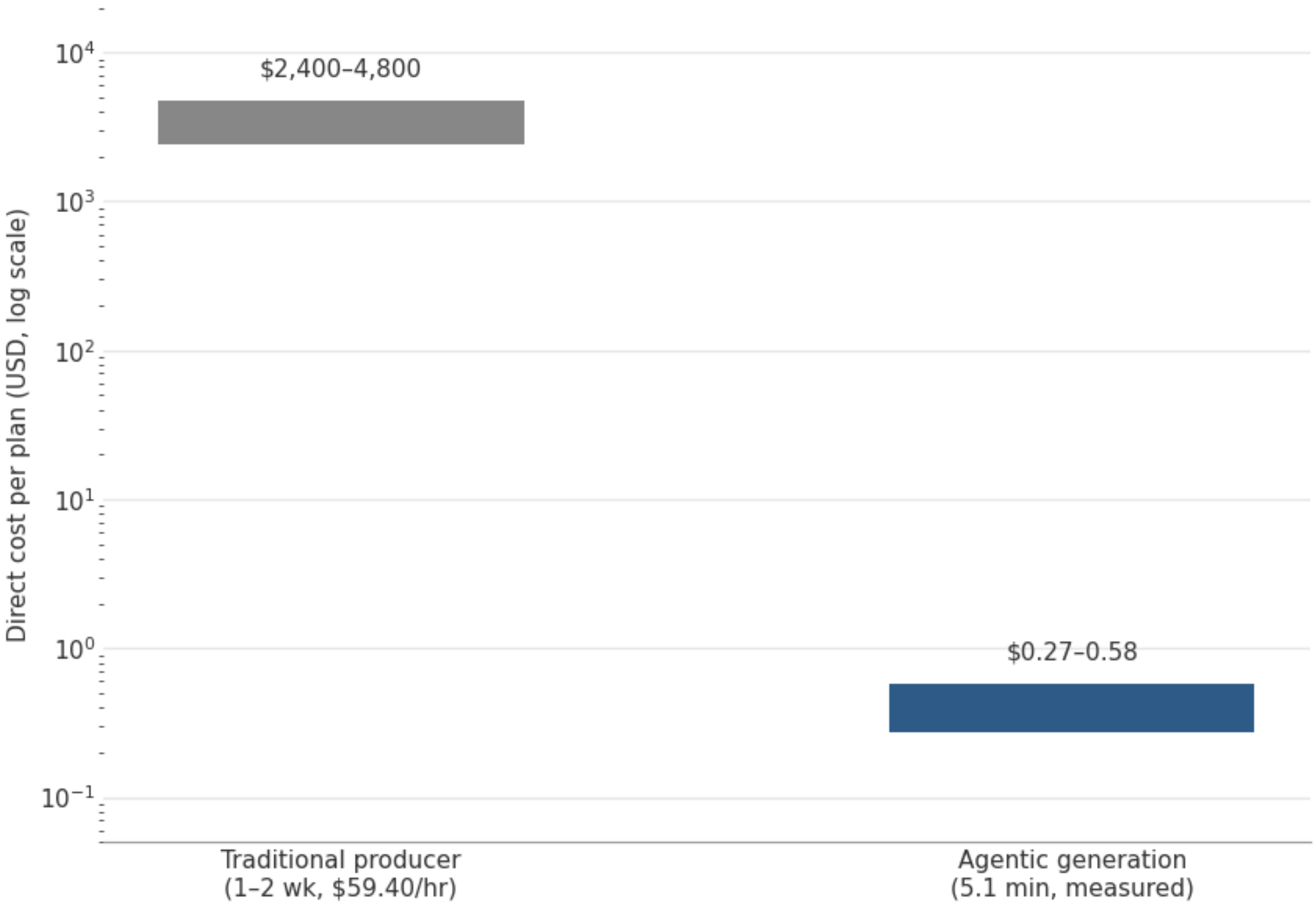

Figure 3. Direct cost per production plan, traditional producer baseline versus measured agentic generation (log scale). The gap spans roughly four orders of magnitude.

The producer-role deliverable, a structured, professionally formatted production plan, repriced from thousands of dollars and weeks of skilled labor to a measured 5.1 minutes and $0.27–0.58, with the dollar cost halving within fourteen months (Table 19). The artifact costing the platform user five minutes and a quarter is the output of a role that most solo and small-team developers cannot afford to staff at $59 per hour [1], which is precisely why, in the first author's direct experience as founder and teacher, they have historically skipped structured production planning altogether.

The comparison underlying Table 6 is a cost-and-time comparison against a labor-market baseline, not a controlled comparison of planning quality or downstream project outcomes. It establishes that the artifact class became radically cheaper; it does not establish that the cheap artifact performs equivalently to the expensive one in use. §4.1.5 and §4.4.3 are this paper's own evidence that the question is open, and the controlled comparison is registered future work (§6).

**The RQ1 answer so far: on dimension D1 (cost) the barrier measurably fell by roughly four orders of magnitude, and on the artifact half of D2 (skill) the deliverable is generable without production-management training. This is the precise sense, per Table 2, in which generative AI has reduced the barrier to entry.**

#### *4.1.4 The regional claim: ARGUED, NOT MEASURED*

RQ1 as posed asks specifically about regions with historically low indie development rates. Per-project creator geography is not available in the current dataset, so this section argues from economics, is labeled accordingly, and the geographic claim is classified ARGUED, NOT MEASURED in Table 2 (D5) and Table 18; measurement is registered future work (§6). The platform's collaborator base spans 38 countries, reported as platform context only.

The access argument: the meaning of a $0.27 production plan is not uniform across geographies. Where a producer's day-rate is a rounding error in a funded studio's budget, the collapse described in §4.1.1 is a convenience. Where that same day-rate exceeds a month's local income, the condition across much of the Global South's emerging development communities, it is the difference between participating in structured production and being excluded from it. The economics are concrete: a producer at the US average of $123,500 per year [1] has never been a hiring option for a self-funded studio in Port Louis, Lagos, or Manila; a $0.27 plan is available to all three. The first author's studio history in Mauritius (§1.2) informs this framing but the argument rests on the arithmetic, not the anecdote. The democratization thesis, if it holds anywhere, holds most where the removed barrier was tallest; whether it does hold there is, on the present evidence, an argued implication awaiting measurement, and the paper claims no more than that.

#### *4.1.5 The boundary of the claim: artifact, not judgment*

The §4.1.1 metrics measure the generation of plans, not the quality of the production judgment embedded in them. A producer's value was never only the artifact; it was the experience encoded in the artifact's choices: what to cut, what to sequence, where projects of this shape historically die. Platform-user feedback in the log period includes the observation that generated tasks were "good but generic" [internal log, paraphrased], which is a user independently discovering this distinction. Observations across the platform's student program (900 enrolled) reflect the same distinction: incoming students can produce a professionally formatted plan on day one, and cannot yet reliably distinguish a good plan from a plausible one. The plan is now free; the judgment is not, and the cost-collapse numbers of §4.1.3 silently substitute the former for the latter unless read with this boundary. Whether commodity planning plus absent judgment nets out positive for inexperienced teams is an empirical question this dataset cannot answer, and the honest prior from decades of software project management is not obviously yes. This boundary recurs in RQ4 (§4.4.3).

### 4.2 RQ2: Market Acceptance

How have games created or assisted by generative AI been accepted in the market? Acceptance has three faces, and they point in different directions: commercial acceptance by players (§4.2.1–§4.2.2), professional acceptance by the industry (§4.2.3), and the structural question of whether the enlarged supply is being absorbed at all (§4.2.4).

#### *4.2.1 Commercial acceptance: the indie expansion*

The five-year release series is the expansion's first exhibit (Table 7, Figure 4).

**Table 7. Steam indie releases by year, 2022–2026. Source: SteamDB, Indie tag, retrieved 2026-07-01 [5]. Cumulative indie catalog: 69,000+ titles.**

| Year | Indie releases | Year-over-year | Context |
|---|---|---|---|
| 2022 | 5,559 | baseline | Pre-generative-AI mass adoption |
| 2023 | 6,006 | +8.0% | ChatGPT/genAI year one |
| 2024 | 8,570 | +42.7% | Inflection; Steam AI-disclosure rules begin (Jan 2024) |
| 2025 | 10,676 | +24.6% | Sustained surge on enlarged base |
| 2026 (through Jul 1) | 4,059 | [partial; naive annualization ~8,100] | Possible plateau, see §4.2.4 |

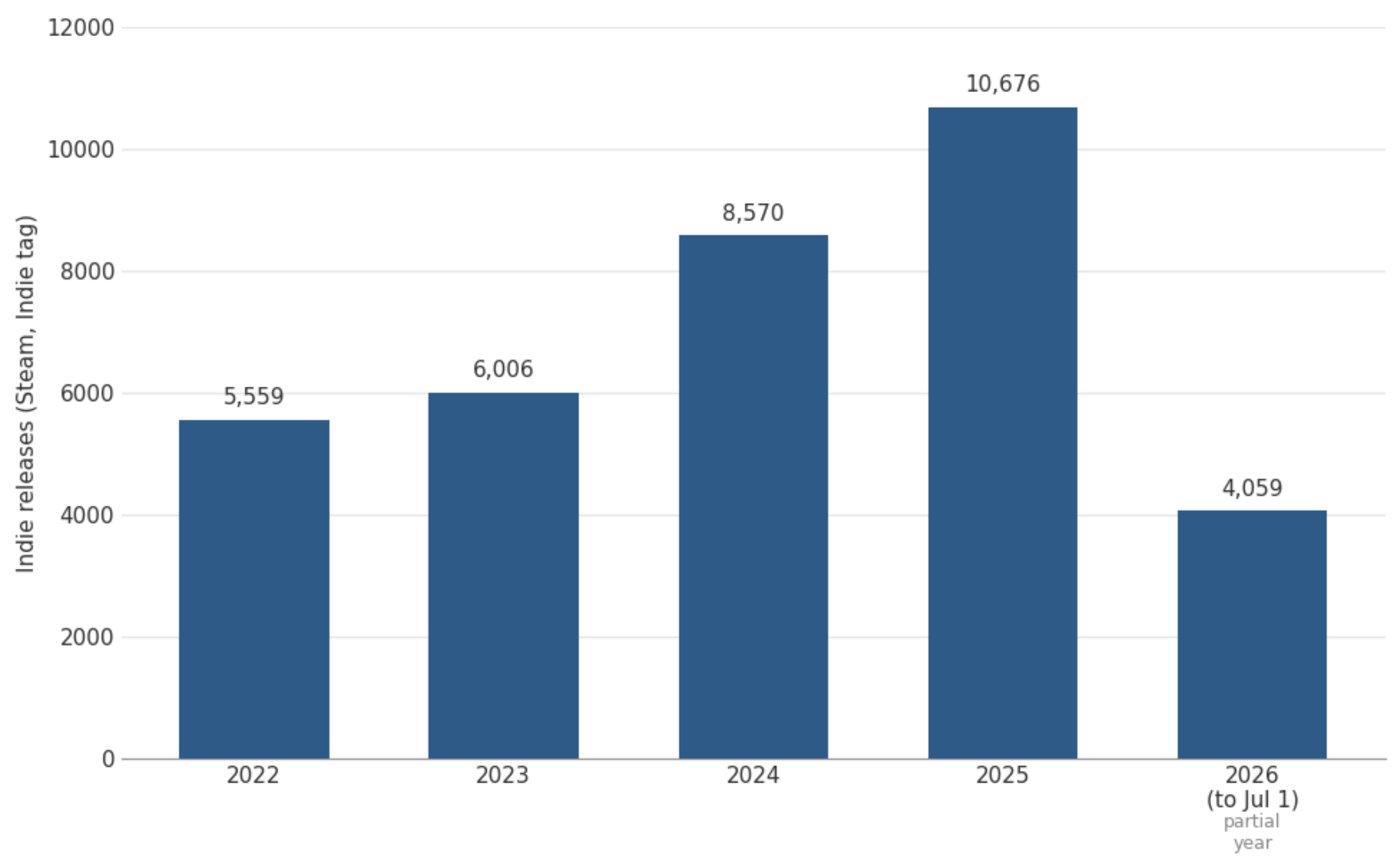

Figure 4. Steam indie releases by year, 2022–2026. The 2024 count marks a +42.7% year-over-year inflection. Source: [5].

The shape matters more than any single number. Growth from 2022 to 2023 was modest, 8%, consistent with the platform's long-run baseline. The 2024 figure represents a 42.7% jump, and 2025 sustained a further 24.6% on the enlarged base. Something changed in the 2023–2024 window, and the timing coincides with the mass availability of generative AI tooling and the emergence of the agentic layer described in §2. Coincidence in timing is not causation, and §3.4's constraints apply; the inflection is located where the wave-three hypothesis predicts it, and its magnitude, a near-doubling of annual output in three years against an 8% baseline, is not subtle. It is also located where several other credible explanations predict it, some of them co-dependent with the tooling story rather than rivals to it; §4.3.3 treats them together, and the paper's causal posture is defined there.

Revenue context sharpens the picture, with one definitional caution: the two most-cited figures use different denominators. Measured against full-game sales revenue only (excluding microtransactions and DLC), independent titles' share on Steam doubled from 24% in 2018 to 48% by 2024, near-parity with AA/AAA for the first time, generating roughly $4 billion in the first nine months of 2024 alone [6]. Measured against Steam's total 2025 revenue of $17.7 billion (a platform record, up 15% year over year), independent games contributed approximately $4.5 billion, over a quarter of everything the platform earned [2].

The unit-sales story is starker still (Table 8).

**Table 8. Best-selling new Steam releases of 2025 by unit volume. Combined: 42 million units. Sources: SteamData Research [30]; corroborated by Alinea Analytics [2] and Valve's official 2025 top-sellers charts [31].**

| Rank | Title | Units sold | Price point | Team profile |
|---|---|---|---|---|
| 1 | R.E.P.O. | 18.4M | under $10 | small indie team |
| 2 | PEAK | 15.7M | $8 | small indie team (Aggro Crab x Landfall) |
| 3 | Schedule I | 8.0M | under $20 | solo/small team, Early Access |

The three best-selling new releases of the year, by volume, were all independent titles priced under $20 from small teams. The expansion is not only in volume but in commercial weight, and it is concentrated exactly where wave-three economics predict: small teams, low price points, viral distribution.

#### *4.2.2 Disclosed-AI titles in the market*

The population most directly relevant to RQ2 as posed, games created or assisted by generative AI, is observable on Steam through the disclosure requirement in effect since January 2024 (Table 9, Figure 5).

**Table 9. Generative-AI disclosure on Steam, 2024–2025. Source: Lambe / Totally Human Media, via Steam's disclosure API [14]. Disclosure is self-reported and loosely enforced; these are floors, not ceilings. Estimated combined gross of disclosed titles: ~$660M. 60% of disclosures concern visual asset generation.**

| Date | Titles with genAI disclosure | Share of total catalog | Share of that year's new releases |
|---|---|---|---|
| Apr 2024 | ~1,000 | 1.1% | n/a (rules began Jan 2024) |
| Jul 2025 | 7,818 | 7% | 20% of 2025 releases (peaking toward 25% in some months) |
| Nov 2025 | 10,258 | 8% | |

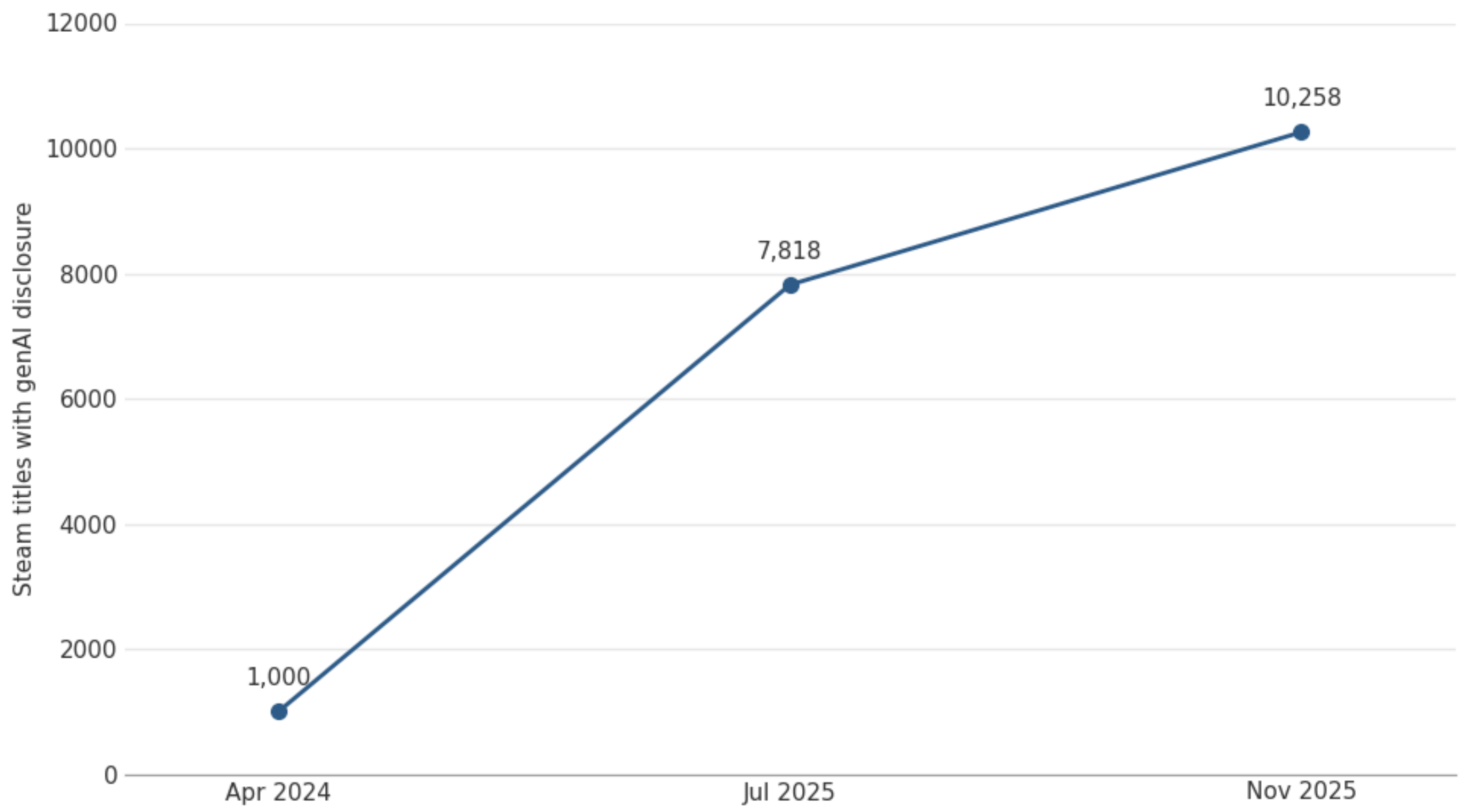

Figure 5. Growth of Steam titles with generative-AI disclosures, April 2024–November 2025; by 2025, roughly 20% of new releases carried a disclosure. Source: [14].

An eightfold increase in disclosed titles in eighteen months, with one in five 2025 releases disclosing generative AI use and an estimated $660M combined gross [14], establishes that AI-assisted titles are being released, purchased, and monetized at scale: the coarsest form of market acceptance. What the disclosure counts cannot establish is reception: whether players accept these titles at the same rate as comparable non-disclosed titles once they meet them.

**Table 10. Player reception of AI-disclosed versus non-disclosed releases: verified-subsample comparison. Disclosure groups from the public replication package of Fabisitooo [27] (Group 1: post-2024 releases with verified store-page disclosure, random subsample of 200 from 424, seed 42; Group 2: pre-2024 releases that added disclosure retroactively, all 80; Group 3: non-disclosed control, all 93); reception metrics from the SteamSpy-derived catalog dataset [28], retrieved 2026-07-07. Review-ratio statistics computed over titles with ≥10 total reviews.**

| Metric | AI-disclosed, post-2024 (n=86) | Retro-labeled, pre-2024 (n=71) | Non-disclosed control (n=54) |
|---|---|---|---|
| Median positive-review ratio | 85.9% (IQR 80.7–93.2) | 84.6% (IQR 78.2–88.9) | 97.8% (IQR 95.7–98.4) |
| Share of titles ≥80% positive | 75.3% | 69.0% | 100% |
| Share of titles ≥95% positive | 14.1% | 7.0% | 79.2% |
| Pooled ratio (review-weighted) | 90.0% | 78.3% | 97.4% |
| Estimated owners ≥20,000 (share of titles) | 40.7% | 73.2% | 72.2% |
| Estimated owners ≥100,000 (share of titles) | 15.1% | 39.4% | 24.1% |
| Median list price | $9.99 | $11.99 | $10.49 |

The result carries one defensible finding and one blocked one. The defensible finding: AI-disclosed 2024–2025 releases in this sample achieve a median positive-review ratio of 85.9%, within Steam's "Very Positive" band, with three-quarters of titles above the 80% threshold; at the coarse level of RQ2, players are accepting disclosed titles at rates typical of the independent catalog, not rejecting them categorically. The blocked one: the 11.9-point gap to the control group (bootstrap 95% CI 8.7–14.4 points; Mann–Whitney $z=-7.70$, $p<0.0001$; robust within the $2–25 paid price band) is not interpretable as a disclosure or quality effect, because the control group's own profile (median 97.8%, every title above 80%) identifies it as a favorably selected comparison set inherited from the source replication package, not a representative draw of non-disclosed releases. The gap therefore bounds the comparison from one side only: disclosed titles underperform a curated set of well-received non-AI titles, while performing at catalog-typical levels in absolute terms. §4.4.2 carries the distributional detail and the registered interpretation rules.

### *4.2.3 Professional acceptance: the sentiment collapse*

Against the commercial expansion, professional acceptance has moved in the opposite direction, and the contraction backdrop explains part of why (Table 11, Figure 6).

**Table 11. Game industry layoffs, 2022–2026. Sources: Ball/Epyllion [3]; gaminglayoffs.com [4]; Noor tracker [20]; Wikipedia synthesis places the cumulative figure in the same ~45,000 range [32].**

| Year | Layoffs | Note |
|---|---|---|
| 2022 | ~8,500 | Wave begins |
| 2023 | ~10,500 | |
| 2024 | ~15,650 | Peak; Embracer collapse contributes ~8,000 across 2023–24 [32] |
| 2025 | ~9,200 | First year-over-year decline (-40%) |
| 2026 (through late June) | ~4,600 est. | 22 studio closures [4] |
| Total 2022–2025 | ~44,000 | 61% North America [3] |

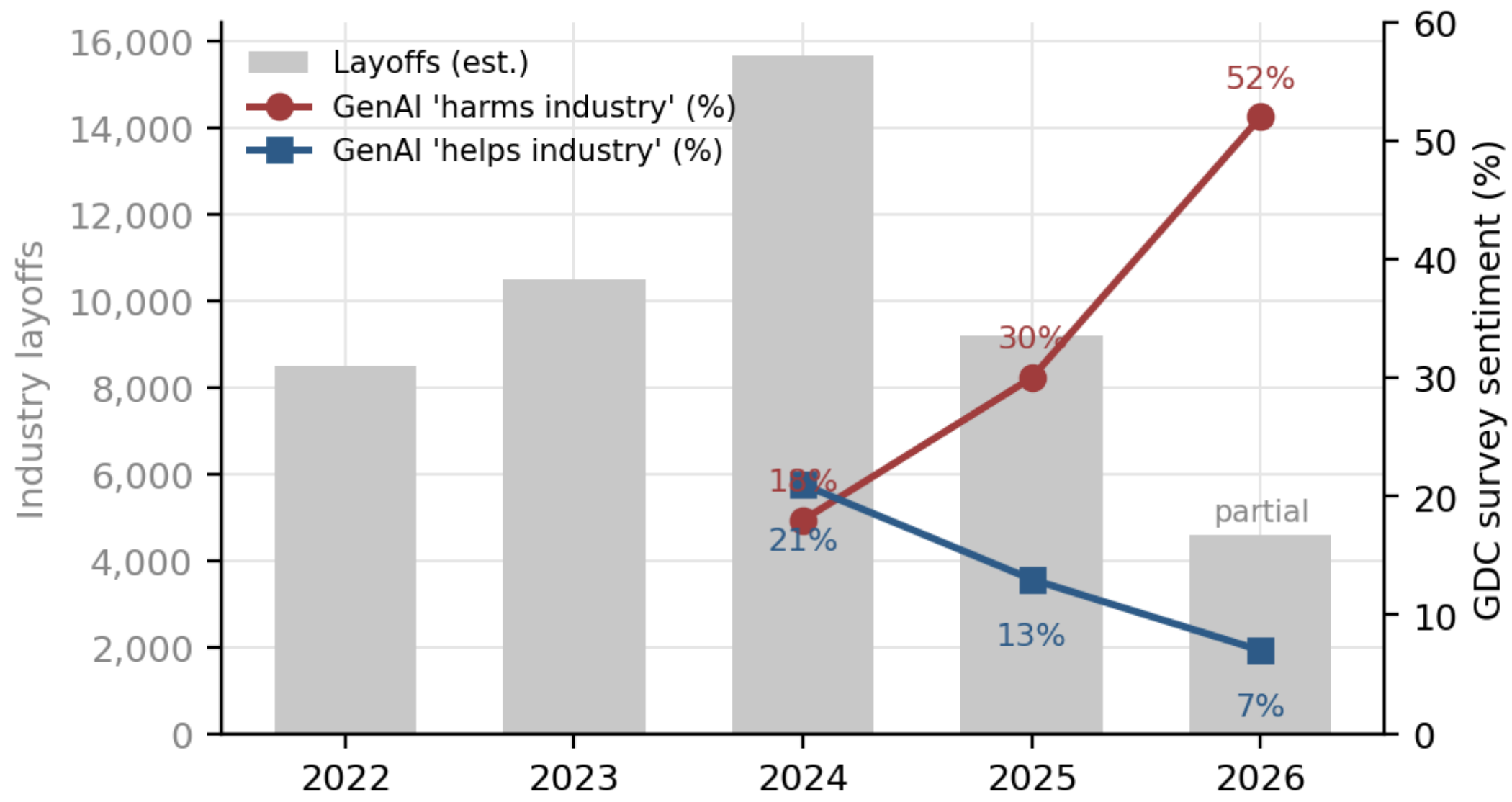

Figure 6. Industry layoffs (bars, left axis) against GDC survey sentiment on generative AI (lines, right axis), 2022–2026. Sources: [3, 4, 7].

The GDC 2026 survey found 28% of respondents globally (33% in the US) had been laid off within the previous two years, and half reported that their current or most recent employer conducted layoffs in the past twelve months [7, p. 13].

**Table 12. GDC survey series: professional sentiment and adoption of generative AI. Source: GDC State of the Game Industry 2025, 2026 [7, 19]; page numbers refer to the 2026 report PDF.**

| Metric | 2024 survey | 2025 survey | 2026 survey |
|---|---|---|---|
| Believe genAI harms the industry | 18% | 30% | 52% (p. 22) |
| Believe genAI helps the industry | 21% | 13% | 7% (p. 22) |
| Actively using AI tools | n/a | n/a | 36% (p. 19) |
| Usage: publishers/support/marketing roles | | | 58% (p. 19) |
| Usage: studio production staff | | | 30% (p. 19) |
| Usage: upper management vs lower-level | | | 47% vs 29% (p. 19) |
| Negative view: visual/technical art | | | 64% (p. 22) |
| Negative view: design/narrative | | | 63% (p. 22) |
| Negative view: programming | | | 59% (p. 22) |
| Primarily self-funded (studio workers and solo devs) | | | 35% (p. 41) |
| Primarily self-funded among solo developers | | | 86% (p. 41) |

The single most striking industry statistic of 2026 is that 52% of surveyed professionals believe generative AI is harming the industry, a figure that has nearly tripled in two years [7, p. 22]: a collapse in professional confidence recorded among the people closest to the work, during the exact window the commercial expansion of §4.2.1 celebrates as an inflection. Practitioner sentiment is not noise; it is a leading indicator produced by thousands of independent observations of studio decision-making from the inside. An analysis that dismisses it as misinformed is an analysis protecting itself. The gradient of that pessimism is itself evidence: it runs steepest among those closest to the craft, visual artists (64%), designers and narrative workers (63%), programmers (59%), and shallowest in the executive suite [7, p. 22].

The adoption data beneath the sentiment contains the pattern most relevant to this paper: AI tool usage is at 36% overall but distributed unevenly, 58% among respondents at publishers, support teams, and marketing/PR firms versus 30% among studio production staff, and 47% in upper management versus 29% in lower-level positions [7, p. 19]. Automation has, so far, penetrated the organizational layer of game development faster than the craft layer, and is most resented exactly where it has penetrated least. Two readings of this asymmetry are available and must both be stated: AI adoption running at 58% in business roles versus 30% in production may mean “AI empowers the organizational layer,” or it may mean the organizational layer is deciding whom to replace. The same datum supports both readings; the data does not choose between them.

The displacement exposure runs through the roles independents hire: the layoff synthesis notes particular exposure among illustrators and other asset-producing roles [32], precisely the freelance categories that independent projects hire from, and that the studied platform's own collaborator marketplace exists to connect. The uncomfortable version of the acceptance story is therefore: wave-three tooling may expand the number of projects while contracting the paid labor each project sustains. The platform log cannot resolve this (it records plans and role identifications, not eventual hiring volumes) and the omission is material. A democratization of project initiation that coincides with a de-professionalization of project labor would be a hollow democratization, and nothing in this paper rules it out.

#### *4.2.4 The saturation counterweight: is the supply being absorbed?*

The structural argument against reading §4.2.1 as unqualified acceptance is the simplest: **production was not the binding constraint.** Discovery was, and remains, untouched. The verified numbers give the squeeze its dimensions, from both sides (Table 13).

**Table 13. Supply up, demand down: the two-sided squeeze.**

| Side | Metric | Value | Source |
|---|---|---|---|
| Supply | Total Steam releases, 2020 | 9,654 | [2] |
| Supply | Total Steam releases, 2025 | 20,000+ | [2] |
| Supply | 2025 releases grossing over $1M | ~300 (~1.5%) | [2] |
| Supply | Indie releases reaching financial viability (2024 cohort) | ~0.5% | [33] |
| Supply | 2025 releases earning effectively nothing | ~half (~9,370) | [2] |
| Demand | Share of global consumer spend, top 8 markets | ~60% | [3] |
| Demand | Gamer participation in those markets vs pre-pandemic | retreating below | [3] |
| Demand | Growth path in mature markets | monetizing a shrinking player base | [3] |

Supply doubled in five years while core-market demand contracted below pre-pandemic levels: the discovery bottleneck is not a friction but a vise. Steam absorbed 25% more indie releases in 2025 into a marketplace whose front-page economics, algorithmic surfacing, and player attention did not grow 25%; every unit of production democratization, in a discovery-constrained market, is partially a transfer of value from creators to the platform that rations visibility, more suppliers competing for the same shelf. The median outcome of a democratized production pipeline, at current discovery economics, is commercial invisibility.

The 2026 mid-year pace is suggestive here: 4,059 releases by July 1 annualizes below the 2025 total. If the full-year figure confirms a plateau or decline, one reading consistent with the saturation case is that the market began pricing in saturation, that the wave-three surge is hitting the discovery wall within two years of starting, far faster than wave two did. [Full-year 2026 figure required before this claim is made or retired.]

The GDC self-funding statistic carries a dark reading in this frame too: 35%+ of studio workers and solo developers primarily self-funding, 86% among solo developers [7, p. 41], describes not only empowered independents but a labor force absorbing its own production risk in a market where the median self-funded title earns effectively nothing (Table 13). Democratized production, concentrated discovery, individualized risk: that is the complete counter-model to the acceptance narrative, and every component of it is consistent with the data of §4.2.1.

**The RQ2 answer so far: AI-assisted titles are being released and monetized at scale (Table 9), the indie category as a whole has been commercially accepted to near revenue parity (§4.2.1), professional acceptance has collapsed in the same window (Table 12), and absorption of the enlarged supply is failing at the median (Table 13). At title level, disclosed releases in the verified subsample are received at catalog-typical rates (median 85.9% positive, Table 10), below a favorably selected non-disclosed comparison set, with the causal reading blocked per §4.4.2.**

## 4.3 RQ3: A New Wave?

Has generative AI begun a new wave of developing, distributing, or adopting games? Section 2 defined the historical signature of a wave: a barrier removed, an output signature, and a forced reorganization of distribution. This section tests the current moment against that signature: convergence between platform and industry measurements (§4.3.1), the capital-substitution mechanism (§4.3.2), the rival explanations (§4.3.3), and the reorganization the wave forces (§4.3.4).

### *4.3.1 Convergence between platform and industry measurements*

**Table 14. Convergence matrix: four pre-registered alignment tests.**

| Test | Platform signal (GH log) | Industry signal | Aligned? |
|---|---|---|---|
| Timing | Operational window Apr 2025 onward | Steam inflection window 2024–2026 (Table 7) | Yes |
| Genre | Top categories: cozy/sim, RPG, roguelike, horror, platformer (Table 5) | Same cluster drives Steam indie surge and 2025 unit-sales leaders (Table 8) | Yes |
| Team scale | Producer deliverable repriced from salary to cents (Table 19) | 35%+ primarily self-funded, 86% among solo devs [7, p. 41] | Yes |
| Scope | One-third of projects are studio operations (§4.1.2) | AI adoption highest in business roles, 58% vs 30% [7, p. 19] | Yes |

Four independent alignments, each registered before results were finalized, each pointing the same direction. Convergence is consistent with the mechanism operating at scale; per §3.4 and §4.3.3, it is not attribution.

### *4.3.2 The capital substitution*

Venture funding for game development fell roughly 85% from its 2021 peak, to about forty deals per quarter by late 2025, during the same window in which 35% of developers became primarily self-funded, 86% among solo developers [7, p. 41], and the cost of producer-grade planning fell to cents (Table 19).

**Table 15. Private funding for game development: peak vs. present. Source: Ball/Epyllion [3].**

| Metric | Pandemic peak | Q4 2025 | Change |
|---|---|---|---|
| Early-stage investment (quarterly) | ~$1.3B (Q3 2021) | ~$200M | ~-85% |
| Pre-seed investment (quarterly) | over $400M (Q1 2022) | under $100M | ~-75% |
| Deal count (quarterly) | 210+ (Q4 2021) | ~40 | ~-81% |

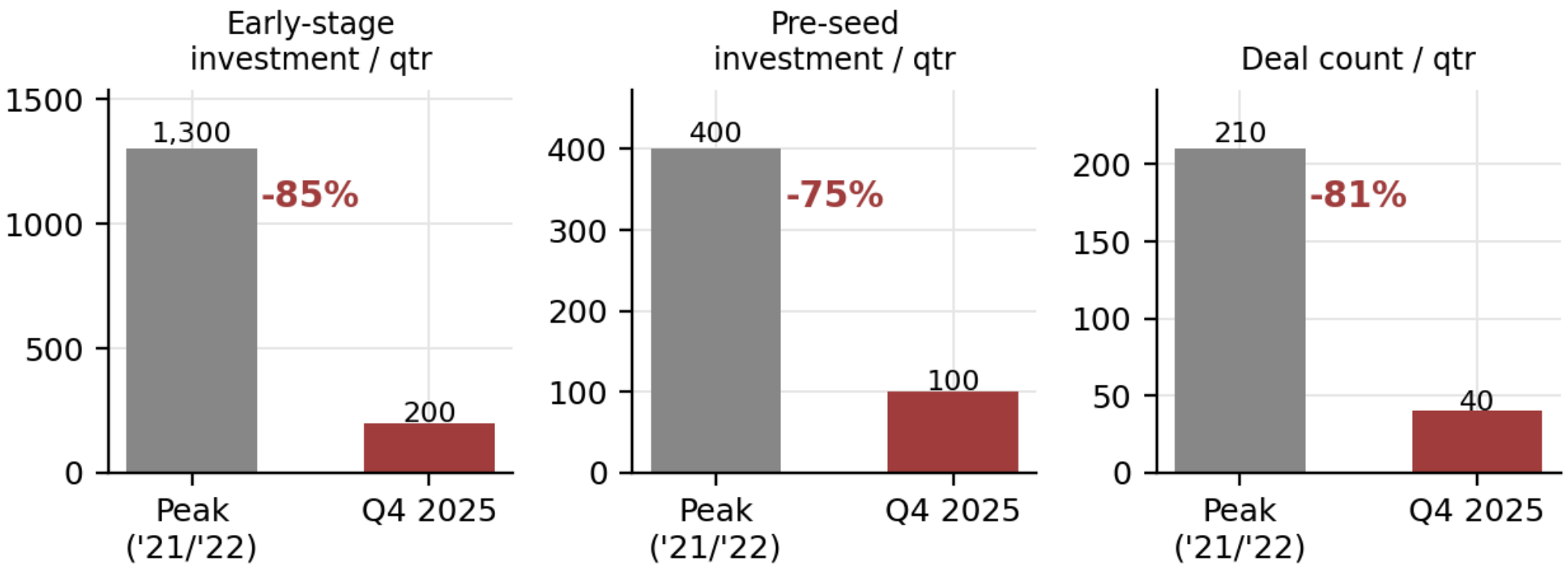

Figure 7. Private funding for game development, pandemic peak versus Q4 2025. Source: [3].

These three series are one story told from three vantage points: the traditional path to a funded production closed, the population of developers proceeding anyway grew, and the tooling that makes proceeding-anyway viable arrived. Self-funding is not only an empowerment story but a residual one; for a large share of independent developers, the traditional capital channel has simply closed. A production-cost collapse arriving at the precise moment the funding channel collapsed is not a convenience but a substitution. If wave three is a wave, this is part of what is driving adoption: not only opportunity, but the absence of the alternative.

#### *4.3.3 Rival and jointly operating explanations*

The timing coincidence of §4.2.1 admits multiple explanations beyond AI tooling, and several are not rivals but co-causes. We name them, with the evidence each carries (Table 16).

**Table 16. Candidate explanations for the 2024 Steam indie inflection.**

| Candidate | Mechanism | Evidence | Assessment |
|---|---|---|---|
| Layoff labor influx | ~44,000 displaced professionals (Table 11), some founding indie teams | GDC 2026: educators report students and displaced workers pursuing indie development and launching their own studios [7, p. 15]; 48% of laid-off respondents had not found another job [7, p. 13] | Plausible co-cause; supplies the population |
| Funding-collapse fragmentation | With VC down ~85% (Table 15), work fragments into smaller, cheaper projects | Direct mechanism from Table 15; 35% self-funding, 86% among solo devs [7, p. 41] | Plausible co-cause; removes the alternative path |
| Engine-market disruption | Unity's 2023 runtime-fee crisis pushed migration; Unreal overtook Unity for the first time (42% vs 30%) [7, p. 28]; Godot gained among newer indies (11%) [7, p. 29] | Timing matches (2023–2024); direction on release volume ambiguous | Possible contributor; direction unclear |
| Steam Direct maturity | The 2017 open door compounds: cohorts of hobbyists mature into shippers | Releases grew every year since 2017; but the pre-2024 trend was ~8%/yr, far below the 42.7% jump | Explains the baseline, not the break |
| Remote-collaboration normalization | Post-pandemic distributed teams lower coordination overhead | Contextual; no direct series | Background condition |
| Generative AI (content) | Asset and code generation lowers content cost | 20% of 2025 releases disclose genAI use, up ~700% YoY [14] | Measured compositional contributor |
| Agentic AI (coordination) | Production-planning cost collapse (this paper) | Direct mechanism measurement at platform level (Tables 19, 6) | Measured mechanism; aggregate contribution not isolated |

Three points of interpretation, registered in advance. First, these candidates are mostly co-dependent rather than competing: a laid-off developer (population) with no funding path (channel closure) still requires the cost collapse (mechanism) to proceed as a self-funded team, and the same GDC report documents displaced workers and students turning to indie formation alongside their tool adoption. Second, this paper's causal claim is correspondingly narrow: we demonstrate the mechanism at micro level (a specific production input repriced by four orders of magnitude) and show four convergences consistent with that mechanism operating at scale; we do not, and cannot with this design, apportion the aggregate inflection among the candidates above. Third, one compositional fact survives all confounders: one in five 2025 releases discloses generative AI in its production [14], so AI-assisted titles observably constitute a large share of the marginal supply regardless of what motivated their creators.

#### *4.3.4 The forced reorganization: toward oversupply and a new distribution paradigm*

If wave three is a wave, the historical pattern (Table 1) predicts its consequence, and it is the paper's central contribution beyond the datasets themselves.

**The mechanism.** When the marginal cost of initiating a professionally structured game project falls toward zero (Table 6) at the same time that traditional capital gatekeeping disappears (Table 15), the classical filter on supply, cost plus gatekeeping, stops operating. Supply is then bounded only by the population of people who want to make games, which wave two already demonstrated is enormous: one million products on itch.io [8], 557,000 jam entries [13], and, in our own log, users planning full studio operations within days of arrival (§4.1.2). The market-structure literature of §2.4 predicts what follows: superstar concentration under near-zero marginal cost [21], against which the long-tail alternative [22] is directly testable with the outcome data of Table 13, and choice-abundant markets whose winners are selected by social dynamics rather than quality gradients [23].

**The forecast.** Under a linear model fit to the four full years 2022–2025 (slope ≈ 1,792 releases/year), annual indie releases reach approximately 12,181 in 2026 and 13,973 in 2027. Under a structural-break model treating 2024 as a regime change and extrapolating the post-break trend (2,106 releases/year), the figures are approximately 12,782 (2026) and 14,888 (2027). The observed 2026 pace tells a different story: 4,059 releases through July 1 annualizes to approximately 8,185, running 33–36% below both projections (Figure 8). Whether this gap reflects early saturation, a data artifact, or seasonal release timing is resolved against the full-year figure at publication time.

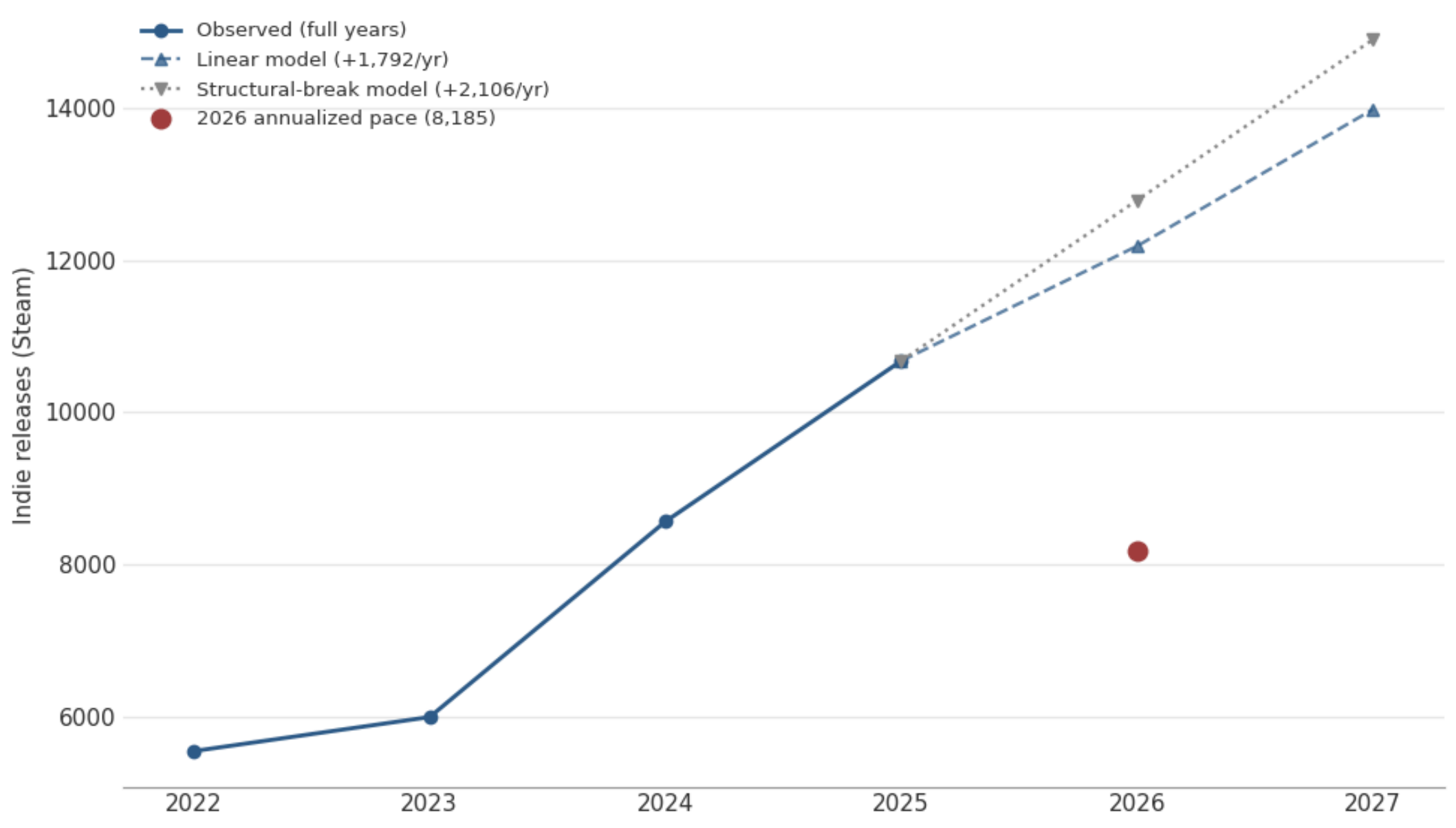

Figure 8. Supply forecast through 2027 under linear and structural-break models, against the observed 2026 annualized pace, which runs 33–36% below both projections.

**The evidence that the reorganization is already beginning.** Figure 9 renders the argument in two panels. Panel A shows both supply projections against the observed 2026 pace, the first quantitative hint that the surplus is already meeting its ceiling. Panel B shows what the 2025 supply actually met in the market: of 20,017 releases, roughly half earned effectively nothing and approximately 300 (1.5%) grossed above $1 million [2]. A market that doubles its supply while halving its median outcome is not growing; it is queuing. Steam releases doubled from 9,654 (2020) to 20,000+ (2025) [2]. One in five 2025 releases already discloses generative AI in production (Table 9) [14]. Meanwhile demand in the eight markets that provide 60% of global spend is contracting below pre-pandemic participation levels [3] (Table 13). Supply doubling into shrinking core-market demand is the standard precondition of surplus: in commodity terms, oversold, an immense excess of choice per unit of player attention and spend.

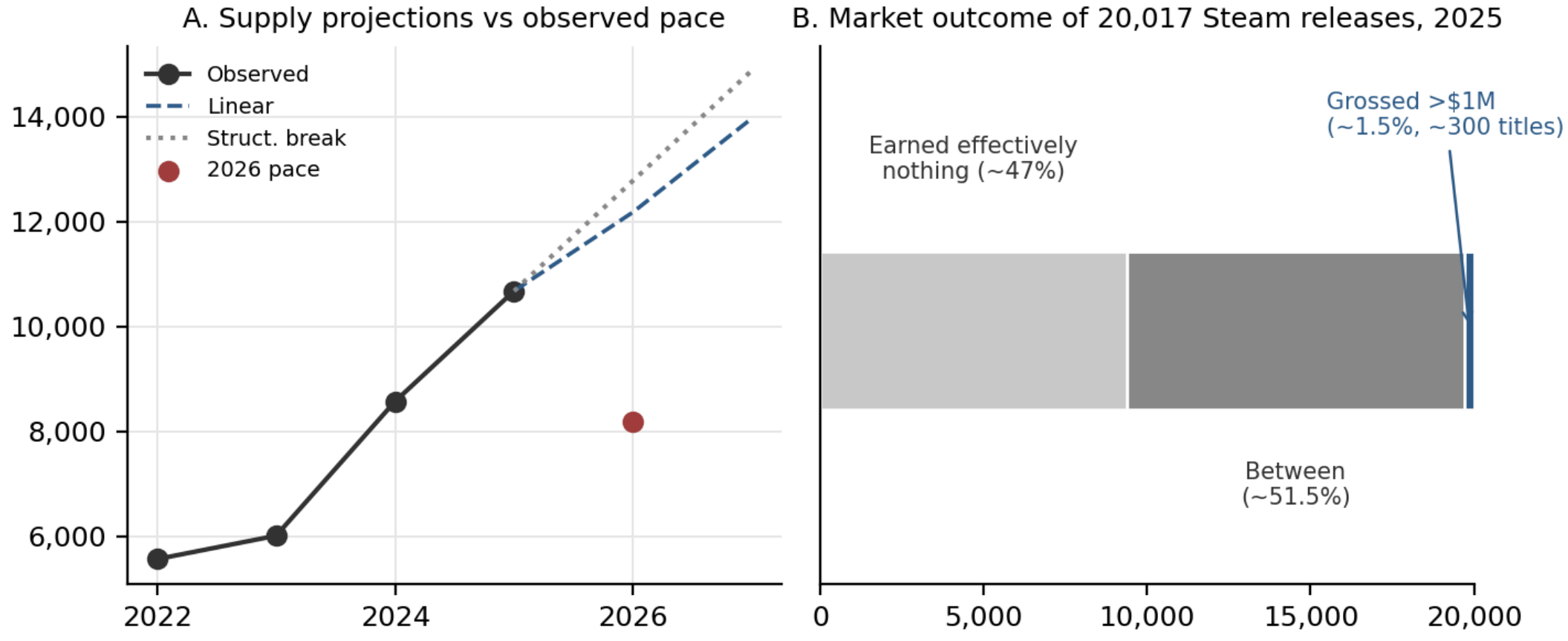

Figure 9. The oversupply argument in two panels. A: supply projections versus observed 2026 pace. B: market outcome of the 20,017 Steam releases of 2025. Sources: [2, 5].

**The counter-scenario, stated and answered.** The strongest objection to the oversupply reading comes from the same source as its demand-side evidence: global demand is not shrinking. Ball's analysis shows worldwide content sales at a record $195.6B, with China contributing 38% of global growth and emerging markets expanding participation [3]. Could global demand growth absorb the supply surplus even as the Major Market 8 contracts? For the supply measured in this paper, the answer the data supports is largely no: emerging-market growth is overwhelmingly domestic and mobile-first, with Chinese consumer spending predominantly on domestic titles [3], while the surplus quantified in Tables 7 and 13 is Western-facing premium PC supply on Steam, whose addressable demand is precisely the contracting Major Market 8. The two curves occupy different markets. If cross-market absorption materializes, emerging-market players adopting Western indie PC titles at scale, the oversupply claim weakens; this is added to the falsification criteria of §5.7.

**The consequence for sellers.** In a surplus market, price and visibility, not production quality, become the binding determinants of commercial outcome. The 2025 data already shows this shape: the year's unit-volume winners were sub-$20 titles propelled by social virality (Table 8), while roughly half of all releases earned effectively nothing and only ~1.5% grossed above $1 million (Table 13) [2, 33]. The industry's own platform-selection behavior confirms where the constraint now sits: developers' top factors in choosing where to release are audience reach (78%) and discoverability (43%), ahead of every technical consideration, in what the survey itself calls an overcrowded marketplace [7, p. 30]. Selling a game is becoming structurally harder not because games got worse but because every game now competes against a functionally unlimited catalog: the player facing 20,000 new titles a year is not underserved, and no marketing budget an indie can afford changes that arithmetic. Our education-program data reflects the identical dynamic at teaching scale: across 900 enrolled students, the reported bottleneck has shifted from completing a playable build to achieving any visibility for a finished one.

**The prediction.** Every prior wave forced a reorganization of how games reach players (Table 1): wave one replaced retail shelves with digital storefronts; wave two replaced curation with the open door and pushed discovery onto algorithms and streamers. Wave three, by removing the last production barrier, breaks the open-door model itself, because an open door into an infinite catalog serves no one. We therefore predict that a new selling and distribution paradigm will emerge, as a structural necessity rather than an innovation choice, with early candidates already visible in the data: virality-native design where the game is its own marketing engine (the R.E.P.O./PEAK pattern of proximity-chat co-op built for clip sharing [2, 30], and marketers' migration to short-video channels, Instagram/Reels 66%, TikTok 64%, YouTube/Shorts 63% [7, p. 45]); community-first and jam-native distribution in the itch.io lineage, where audience precedes product; curation and trust layers re-emerging above the storefront (SteamDB's automatic AI-content tag [14] is an early, crude example of players demanding new filters); demand-side aggregation through subscription, UGC platforms, and creator ecosystems capturing attention upstream of stores [3]; and verified-quality signals, of which registered, methodologically transparent development research, the practice this paper itself models, is one candidate trust primitive. Which of these becomes the wave's storefront is an open empirical question; that one of them must is, we argue, the direct implication of Tables 6, 7, 9, 13, and 15 read together. The next democratization wave, if there is one, is a discovery wave, and nothing in current AI tooling addresses it.

**The RQ3 answer: on developing, yes, in the bounded sense of Tables 2 and 14, a measured cost collapse in a core production input with four convergences at industry scale, though the aggregate inflection is not causally apportioned (Table 16). On distributing and adopting, the wave's distribution reorganization is predicted, with early candidates named and falsification criteria registered, not yet demonstrated.**

## 4.4 RQ4: Quality

What is the quality of the games that have been created with generative AI? The platform log measures the generation of production plans, not the quality of shipped games; no title in the instrumented sample has a verified public release [status check registered as an open item]. This question is therefore answered from marketplace outcome data, with the definitional boundary stated first.

### *4.4.1 The definitional boundary*

Steam's disclosure requirement, in effect since January 2024, records generative-AI use anywhere in a title's production; 60% of disclosures concern visual asset generation [14]. No public data source isolates games authored end-to-end by generative AI. The measurable population is therefore "titles disclosing generative-AI use in production," and every finding in this section is scoped to that population. "Quality of games created by GenAI," as the question is posed, is answerable today only in this proxied form, and the paper says so rather than pretending otherwise. Disclosure is additionally self-reported and loosely enforced [14], making the disclosed group a floor: some AI-assisted titles sit unlabeled in the control group, which biases any measured difference toward zero. Findings of difference are therefore conservative; findings of no difference are ambiguous between genuine parity and contamination.

### *4.4.2 Quality signals: disclosed versus non-disclosed releases*

The registered design of §3.3 is executed here in a verified-subsample form (the deviation and its reasons are documented in §3.3): disclosure status from an independent replication package [27], reception metrics from a SteamSpy-derived catalog dataset [28], positive-review ratio as the primary quality signal with a 10-review floor. Playtime and Metacritic fields are unpopulated in the retrieved distribution and are deferred to the full-census upgrade; ownership tiers substitute as the behavioral-acceptance signal (Table 10).

**Table 17. Quality-signal distribution, AI-disclosed versus non-disclosed releases (verified subsample; groups and sources per Table 10). Statistical comparison by Mann–Whitney U (tie-corrected normal approximation) on title-level review ratios; CI by bootstrap over medians (B=20,000).**

| Comparison | Median difference | 95% CI | Test |
|---|---|---|---|
| Disclosed post-2024 vs. control | -11.9 pp | (-14.4, -8.7) | z=-7.70, p<0.0001 |
| Retro-labeled vs. control | -13.2 pp | (-16.0, -11.5) | z=-8.66, p<0.0001 |
| Disclosed vs. control, $2–25 price band | 86.7% vs. 97.8% | n=72 vs. 47 | z=-7.21, p<0.0001 |

Applying the interpretation rules registered in advance of results: the lower ratio among disclosed titles is consistent with lower intrinsic quality, with player disclosure-aversion (the label itself depressing ratings), and, in this execution, with a third mechanism that dominates both: favorable selection of the control set, whose distribution (median 97.8%, minimum above 80%) is characteristic of curated well-received titles rather than of the release population. The design cannot separate these, and the control-selection issue means the measured gap is an upper bound on any true disclosure effect. What survives the confounds is distributional: disclosed titles cluster in the 80–93% band, the ordinary range of competently received independent releases, with a thin tail above 95% (14.1% of disclosed titles versus 79.2% of the curated controls). If the plan-level prediction of §4.4.3 (compression toward a competent middle rather than uniform degradation) generalizes to shipped games, this is what it would look like: a solid median, a thin excellence tail. The pattern is consistent with that prediction; the sample cannot confirm it, and the full-census matched design remains the registered upgrade.

### *4.4.3 Qualitative signals from the production layer*

The platform log contributes one bounded quality observation, at the plan level rather than the game level: user feedback in the log period describes generated tasks as "good but generic" [internal log, paraphrased], and observations across the platform's student program (900 enrolled) reflect the same distinction: incoming students can produce a professionally formatted plan on day one, and cannot yet reliably distinguish a good plan from a plausible one (§4.1.5). If the pattern generalizes from plans to games, the quality question of RQ4 becomes a judgment question: commodity generation raises the floor of formal competence while leaving the ceiling, the taste and experience that separate a plausible artifact from a good one, untouched. Table 17 and the distribution of Table 10 are consistent with that pattern in shipped-game reception: a solid competent median with a thin excellence tail, though the sample constraints of §4.4.2 prevent confirmation.

### *4.4.4 Quality of the field: the homogenization risk*

Quality has a second sense at the level of the medium rather than the title: the diversity of what gets made. The strongest evidence of risk here comes from this study's own dataset. The log contains conspicuous clustering: multiple near-identical cozy-gardening-and-trading prototypes created within the same weeks; the same project brief ("3D Time-Manipulation Adventure Game") planned repeatedly by different, unrelated users; recurring genre-template structures across unrelated accounts. Some of this clustering has benign explanations: a platform game jam with a shared theme ran during the window, and onboarding flows encourage template starts. But the concerning reading is fully consistent with the same observations: when generation cost approaches zero, the marginal project gravitates toward the template, and the design space compresses even as the project count expands.

Wave one demonstrably widened the range of games that could exist commercially. Wave three has so far demonstrated only that it widens the count. If the count rises while the range narrows, the democratization thesis fails on its own terms: more people making the same game is not a democratized medium. We flag this, deliberately, as a decisive open question of the wave, and one this dataset is positioned to answer with more data: the upgrade path (§5.7), deduplication, developer-assigned tags, and similarity measurement across the full corpus, converts this section's anecdote into a testable claim.

**The RQ4 answer so far: at title level, disclosed releases show catalog-typical reception at the median (85.9% positive) with a thin tail above 95%, a gap to a favorably selected control that bounds any true effect from one side only (Tables 10, 17); at plan level, formal competence is commoditized while evaluative judgment is not; at field level, homogenization risk is documented in this study's own data and registered as the wave's decisive open question.**

# 5 DISCUSSION

The four questions resolve not by splitting differences but by noticing that the popular debate has collapsed distinct variables into one. We state the results as results, each mapped to the dimensions of Table 2.

## 5.1 Results by research question

**RQ1.** The production-planning barrier fell by roughly four orders of magnitude in cost and roughly three in time (Tables 19, 6): dimension D1 demonstrated, and the artifact half of D2. The regional extension of the claim (D5) is argued from cost arithmetic and remains ARGUED, NOT MEASURED. What fell was the artifact, not the judgment (§4.1.5).

**RQ2.** Acceptance is split three ways: commercial acceptance of the indie category approached revenue parity (§4.2.1) and AI-disclosed titles are monetizing at scale (Table 9); professional acceptance collapsed across three consecutive surveys (Table 12); and absorption is failing at the median, with roughly half of releases earning effectively nothing (Table 13). At title level, disclosed releases are received at catalog-typical rates (median 85.9% positive, Table 10), with the comparison to non-disclosed titles bounded but not causally identified (§4.4.2).

**RQ3.** The current moment fits the historical wave signature in the bounded sense: mechanism measured at micro level, four registered convergences at macro level (Table 14), aggregate attribution not apportioned among jointly operating causes (Table 16), and the wave's forced distribution reorganization predicted with registered falsification criteria (§4.3.4). Democratization is also partly a substitution for collapsed capital (Table 15): the traditional path closed, the population proceeding anyway grew, and the tooling that makes proceeding-anyway viable arrived.

**RQ4.** At title level, the verified-subsample comparison finds a competent median (85.9% positive) with a thin excellence tail among disclosed releases, and an 11.9-point gap to a favorably selected control that upper-bounds any true disclosure effect (Tables 10, 17); at plan level, commodity generation raises formal competence while leaving evaluative judgment scarce; at field level, this study's own data documents homogenization risk (§4.4.4), evidence bearing against D6.

## 5.2 Unresolved, honestly: the layoff question

On AI's causal role in the AAA contraction, both popular narratives outrun the evidence. The displacement case rests on sentiment and role-exposure patterns; the exculpatory case rests on companies' own stated reasons, in which automation appears at just 6% [7, p. 16]; both are confounded by the post-pandemic correction, and the public record cannot separate them. This paper's contribution to that debate is deflationary: the measurable action of AI in games is at the production-barrier margin, where its effect is large, quantified, and expansionary, whatever its unmeasured role in AAA headcount may be. The two questions should stop being argued as one.

## 5.3 The access dividend (D5)

Whatever the net industry-wide effect, the distribution of the production dividend is progressive: it is largest where the traditional barrier was tallest relative to local income. Should future geographic measurement across the platform's 38-country base confirm Global South participation, this would be the democratization thesis's most defensible surviving form: not “AI is good for the industry,” but “AI-repriced production disproportionately benefits those the prior cost structure excluded.” Because single trajectories, including the first author's own (§1.2), prove nothing on their own, the paper defers this claim to measurement rather than asserting it.

## 5.4 The classroom signal

From the educator's vantage, structured practitioner observation, explicitly not a formal study, the synthesis has a concrete curricular consequence. Students at the intersection of the first author's teaching (University of Silicon Valley) and the platform's student program (900 enrolled; free access via .edu) now arrive at production planning as a day-one commodity rather than a taught-late professional skill. The GDC 2026 survey's education supplement sharpens the picture from both directions: educators surveyed held more favorable views of generative AI than their own students (58% of educators use the tools versus 39% of students; 53% of educators call the impact mixed while 57% of students call it negative, with the report's own small-sample caveat) [7, p. 23], while students cited AI-led displacement, alongside the scarcity of entry-level roles and competition from laid-off veterans, among their chief anxieties about entering the industry, with 74% concerned about their job prospects [7, p. 15]. The RQ1/RQ4 boundary dictates what replaces planning in the curriculum: if the artifact is free and the judgment is scarce, game design education must shift from producing plans to evaluating, editing, and rejecting them, judgment about structure, taught explicitly, against AI-generated baselines. And §4.3.4 dictates the second shift: in a surplus market, distribution literacy (community building, virality-native design, discovery economics) moves from an elective afterthought to a core discipline. A formal cohort study (users versus non-users of agentic planning tools) is specified as future work.

### 5.5 Synthesis statement

AI tooling demonstrably collapsed the time and dollar cost of professional-grade production practice, measured at 5.1 minutes and $0.27–0.58 per plan against a $2,400–4,800 traditional baseline, during the same window in which independent output inflected sharply upward, with four registered convergences linking the two scales. The same evidence base shows the gains concentrated in artifact production rather than production judgment, the output expansion unproven as a diversity expansion, and a market exhibiting the preconditions of structural oversupply: supply doubled while core-market demand contracted, with roughly half of new releases earning nothing. The democratization of wave three is real, partial, dimensionally bounded (Table 2), and, like the two waves before it, a relocation of scarcity rather than an abolition of it: this time from production to attention. When making games has no barrier, selling them becomes the barrier, and the distribution paradigm that resolves this, whatever form it takes, will define the next era of the medium.

### 5.6 Claims and evidence status

For review and replication, the paper's claims are consolidated by evidential status in Table 18.

**Table 18. The paper's claims, by evidential status.**

| Claim | Status | Basis |
|---|---|---|
| The production-planning deliverable repriced by ~4 orders of magnitude (D1) | Demonstrated | Direct measurement, Tables 19 and 6 |
| The artifact is generable without production-management training (D2, artifact half) | Demonstrated | System behavior, Table 19 |
| Platform patterns converge with industry trends | Demonstrated (as convergence) | Four registered tests, Table 14 |
| AI tooling contributed to the 2024–2026 output inflection (D3) | Supported, not isolated | Mechanism + convergence + 20% disclosure share; confounders in Table 16 |
| Barrier reduction extends to low-participation regions (D5) | Argued, not measured | Cost arithmetic, §4.1.4; measurement registered future work |
| Democratization is partly substitution for collapsed capital | Supported | Co-movement of Tables 12, 15, 19 |
| The market shows preconditions of structural oversupply | Supported, falsifiable | Tables 7, 13; criteria in §5.7 |
| A new distribution paradigm will emerge | Predicted | §4.3.4 argument from Table 1 pattern + §2.4 theory [21–23] |
| AI-disclosed titles are commercially accepted at scale (coarse) | Supported | Table 9; ~$660M disclosed-title gross [14] |
| Disclosed titles receive catalog-typical reception (median 85.9% positive) | Measured (verified subsample) | Tables 10, 17 |
| Disclosed titles receive worse reception than non-disclosed titles | Not claimed; bounded only | Control favorably selected; gap is an upper bound (§4.4.2) |
| Judgment quality of AI plans equals producer-authored plans | Not claimed | Registered future work |
| AI causes AAA layoffs | Not claimed | §4.2.3, §5.2; 6% company-cited rate [7, p. 16] |
| Diversity of output is expanding (D6) | Not claimed; evidence of risk | §4.4.4 |
| Participant commercial outcomes improved (D7) | Not claimed; evidence against | Table 13 |

### 5.7 Limitations

Single-platform sample; self-selected users (those who chose an AI tool cannot represent those who did not). Fourteen-month log window against a longer platform history; the log is a sample of a corpus roughly three times larger [referent under confirmation]. Genre classification is keyword-derived from titles rather than structured metadata; the registered upgrade path is deduplication plus developer-assigned tags on the full corpus. Convergence is not causation, and no counterfactual exists for what these developers would have done without the tooling; §4.3.3 names the jointly operating explanations and the paper's causal claims are bounded to the measured mechanism. The log measures planning, not releases: the conversion rate from generated plans to shipped games is unmeasured here and is the single most important missing number in the platform study. Approximately 27% of creation events are internal accounts, excluded from adoption claims but present in system metrics. Geographic analysis is limited to the labeled economic argument of §4.1.4; RQ1's regional sub-claim is answered as ARGUED, NOT MEASURED.

Education observations are practitioner observation, not controlled study.

itch.io trend data does not exist publicly, leaving Steam as the sole longitudinal storefront series, a series that likely under-represents the most informal tier of output, making the supply-side figures conservative. The revenue-share statistics of §4.2.1 derive from two analysts using different denominators (full-game sales versus total platform revenue), reported separately and labeled as such; neither is an official Valve figure.

The title-level analyses of §4.2.2 and §4.4.2 carry their own limitations. Stated in advance of results: disclosure is self-reported and loosely enforced, so any control group is contaminated with unlabeled AI-assisted titles, biasing measured differences toward zero; "disclosed" is not "created by," and 60% of disclosures concern visual assets only [14]; and review ratios confound intrinsic quality with disclosure-aversion among players. Arising from the executed verified-subsample form (§3.3): the inherited control group is favorably selected (median 97.8% positive, all titles above 80%), so the disclosed-versus-control gap is an upper bound on any true effect and is not causally identified; the metrics dataset undercovers very recent and very small titles (match rates 43–89% across groups), biasing all groups toward visibility and excluding the smallest releases whose reception the saturation analysis of §4.2.4 concerns most; both source groups were constructed conditional on review volume, removing the "share reaching the review floor" metric; and playtime and Metacritic signals were unavailable in the retrieved distribution. These constraints are why the paper claims catalog-typical reception for disclosed titles and declines to claim a reception deficit; both claims are stated separately in Table 18.

The oversupply prediction of §4.3.4 is an argued implication of measured trends, not itself a measured outcome; it is stated falsifiably (full-year 2026 and 2027 release, revenue-concentration, and participation figures will test it, as would evidence of large-scale emerging-market absorption of Western indie PC supply). No comparison of plan quality between AI-generated and producer-authored artifacts is made here; Table 6 compares cost and time only, and the controlled comparison is registered future work. The first author co-founded the platform studied; the study registration is the primary control, the platform-independent title-level analysis is a second, and the adverse findings reported in §4.1.5, §4.2.3, §4.2.4, §4.4.3, and §4.4.4 are the demonstration that the commitment to unfavorable evidence was honored.

## 6 FUTURE WORK

The following items, in order of decisiveness, are registered as future work building directly on the results above.

Future work, in order of decisiveness: the full-census execution of the title-level analyses (upgrading the verified subsample of Tables 10 and 17 to the complete disclosure population with representative matched controls); a controlled comparison of traditional versus AI-assisted planning (matched projects, or blinded expert evaluation of producer-authored versus L1-generated plans, on structural quality, feasibility, risk coverage, and plan-to-milestone survival); plans-to-releases conversion tracking; deduplicated, tag-based similarity measurement across the full corpus (the homogenization test); full-year 2026 and 2027 supply, revenue-concentration, and participation figures (the oversupply test); cross-platform replication; per-project geographic analysis of the access dividend; and a formal student-cohort study of judgment formation under commodity planning.

A further line of work follows directly from the oversupply finding of §4.3.4 rather than from a measurement gap in this paper. If the market is entering a period of structural oversupply, two questions become immediate rather than speculative. The first is the initial market impact of that oversupply: how visibility, discovery, and revenue concentrate or fragment in the first one to two years after the inflection, and whether the early 2026 pace reported in §4.3.4 is a leading indicator of that impact or a transient artifact. The second is the player-side response: as the catalog grows larger and less differentiated, how players solve the problem of finding titles that match their play styles, whether through platform recommendation systems, curator- and creator-driven discovery, or genre- and community-based filtering, and whether that discovery layer becomes the next bottleneck once the production barrier measured here has fallen. Both are registered as direct extensions of the market-structure argument of §4.3.4, not as claims this paper makes.

## 7 CONCLUSION

RQ1: the barrier to a core production input measurably collapsed, professional-format plans of ~16 epics and ~59 stories generated in ~5 minutes at $0.27–0.58 against a traditional baseline of $2,400–4,800 and one to two weeks (Tables 19, 6), applied two-thirds to game design and one-third to studio operations (Table 5); the regional extension of the claim is argued from cost arithmetic and awaits measurement. RQ2: the indie category approached revenue parity and AI-disclosed titles monetize at scale (Tables 7–9), while professional acceptance collapsed (Table 12) and median absorption failed (Table 13); at title level, disclosed releases are received at catalog-typical rates (median 85.9% positive; Table 10). RQ3: four registered convergence tests align (Table 14), the capital-substitution mechanism co-moves (Table 15), aggregate attribution remains unapportioned among jointly operating causes (Table 16), and the wave's predicted consequence, a forced distribution reorganization under structural oversupply, is stated falsifiably (§4.3.4). RQ4: the verified-subsample quality comparison finds a competent median with a thin excellence tail among disclosed releases and no causally identifiable deficit (Tables 10, 17); the production-layer evidence indicates commoditized formal competence with scarce evaluative judgment, and the study's own data documents homogenization risk at field level (§4.4.4).

The paper's lens contribution is the three-wave framing plus the dimensional operationalization of democratization (Table 2), under which the current moment is a structural phase with two precedents, a predictable signature, and a predictable consequence. Its empirical contributions are the first longitudinal, quantitative, project-level dataset of agentic production tooling in independent game development, and the first matched title-level reception comparison of AI-disclosed releases, executed here as a verified subsample (§4.2.2, §4.4.2). Its forward contribution is the oversupply thesis: when making games has no barriers, the barrier moves to being played, and the market's response to that relocation, not the production tooling that caused it, will define who thrives in the next decade of independent games. Its within-section dialectical structure is offered as a template for founder-authored platform research: each claim is only as credible as the counter-evidence it is presented beside.

# A INSTRUMENTED PLAN GENERATIONS

**Table 19. Instrumented plan generations, per project (n=40). † internal/team/test account (retained for system metrics per §4.2; 13/40 rows). Type source codes: (k) registered keyword match on title; (l) corroborated by related log entries; (t) descriptive inference from title, annotation only; (u) not inferable from title. n/l = not logged (telemetry fields added incrementally). Tasks column omitted: task decomposition occurs post-planning and was 0 at generation for all rows.**

| Project | Type (source) | Epics | Stories | Skills | Tools | Time (s) | Cost |
|---|---|---|---|---|---|---|---|
| P. Neptune | Unspecified, largest plan in sample (u) | 33 | 122 | 2 | 5 | 493 | $1.20 |
| Epic Quest | Fantasy RPG (k) | 26 | 99 | 3 | 5 | 398 | $0.95 |
| X-Com Cyberpunk | Tactical strategy (X-COM-like) (k) | 25 | 99 | 3 | 3 | 333 | $0.94 |
| Shadow Realm | Horror-action roguelike (l) | 24 | 89 | 3 | 5 | 348 | $0.87 |
| Digital Warriors | Action RPG (k) | 24 | 88 | 3 | 5 | 351 | $0.84 |
| Clock Out at 2 Plan | 2D horror roguelite (vertical slice) (l) | 23 | 81 | 4 | 4 | 321 | n/l |
| (Pending) | Untitled at generation (u) | 22 | 84 | 4 | 5 | 371 | $0.80 |
| Pokebiri | Creature-collection (title-inferred) (t) | 22 | 76 | 6 | 5 | 345 | $0.76 |
| Digital Warriors † | Action RPG (k) | 21 | 73 | n/l | n/l | 481 | n/l |
| Rents Cheap | Unspecified (u) | 21 | 82 | 3 | 5 | 321 | $0.79 |
| Lucifer’s Gauntlet | Action game (core prototype + GDD) (l) | 20 | 74 | n/l | n/l | 362 | n/l |
| Epic Quest † | Fantasy RPG (k) | 20 | 72 | 3 | 5 | 336 | $0.69 |
| The game of games | Unspecified (u) | 20 | 78 | 4 | 5 | 311 | $0.75 |
| PewPew | Arcade shooter (title-inferred) (t) | 20 | 77 | 4 | 3 | 268 | $0.72 |
| PewPew | Arcade shooter (title-inferred) (t) | 20 | 62 | 4 | 4 | 294 | $0.66 |
| Shadow Realm † | Horror-action roguelike (l) | 19 | 74 | 7 | 5 | 340 | $0.72 |
| Modern God’s † | Unspecified (u) | 19 | 76 | 4 | 4 | 313 | $0.76 |
| Super Mario Splatter † | Platformer (parody, internal test) (k) | 18 | 67 | 2 | 5 | 321 | $0.65 |
| Test | Internal/user system test (u) | 18 | 67 | 4 | 5 | 303 | $0.69 |
| Funny Horror Tetris † | Horror puzzle (internal test) (k) | 17 | 58 | n/l | n/l | 310 | n/l |
| Test | Internal/user system test (u) | 17 | 61 | 3 | 5 | 372 | $0.63 |
| Strings and Nails | String-art puzzle (l) | 15 | 60 | 4 | 4 | 274 | $0.58 |
| Neon Dreams † | Unspecified (u) | 14 | 58 | 3 | 5 | 336 | $0.55 |
| Maze of Spiral | Maze survival (k) | 14 | 53 | 4 | 5 | 299 | $0.52 |
| DESYNC | Unspecified (u) | 14 | 55 | 4 | 4 | 242 | $0.53 |
| Afro-Future: Rising † | Afro-futurist project (internal) (t) | 13 | 57 | n/l | n/l | 346 | n/l |
| Duck Hunt | Arcade shooter (title-inferred) (t) | 13 | 49 | 2 | 3 | 278 | $0.49 |
| Test † | Internal/user system test (u) | 11 | 43 | 4 | 4 | 245 | n/l |
| Void Miner † | Incremental asteroids roguelite (l) | 10 | 38 | 3 | 4 | 242 | $0.38 |
| Strike Zombie | Zombie action (k) | 10 | 36 | 3 | 4 | 277 | $0.37 |
| The Rise of the Phoenix | Fantasy (k) | 9 | 30 | 3 | 4 | 264 | $0.34 |
| Circuit Bloom | Unspecified (u) | 9 | 29 | 4 | 5 | 249 | $0.32 |
| Cafe Connisseurs | Cafe management / cozy sim (title-inferred) (t) | 9 | 38 | 3 | 4 | 243 | $0.39 |
| Duped † | Unspecified (internal test) (u) | 8 | 24 | 3 | 3 | 222 | $0.29 |
| DontLook! | Horror (title-inferred) (t) | 8 | 28 | 3 | 3 | 254 | $0.33 |
| Maze Solitaire | Maze card-puzzle (k) | 8 | 33 | 2 | 5 | 237 | $0.34 |
| Kingdom Hearts 4 † | Action RPG (existing-IP internal test) (k) | 5 | 22 | 0 | 3 | 224 | $0.23 |
| WAR RATS: The Rat em Up † | Beat-em-up (internal test) (t) | 5 | 18 | 0 | 3 | 419 | $0.21 |
| Portfolio | Portfolio site (non-game) (t) | 5 | 20 | 3 | 3 | 210 | $0.24 |
| Escape Of The Medic | Escape adventure (title-inferred) (t) | 5 | 16 | 1 | 2 | 194 | $0.20 |

Summary: epics mean 15.8 (median 17, range 5–33); stories mean 59.1 (median 60, range 16–122); stories per epic mean 3.7; skills identified mean 3.2 (range 0–7, n=36); tools identified mean 4.2 (range 2–5, n=36); cost per plan mean $0.58 (n=34); cost per story mean $0.010.

**Table 20. Plan complexity by game type (registered keyword classification, instrumented subsample n=40, multi-label). A project counts in every matched category. n/l = not logged for the only project(s) in that cell. Small per-cell n; read as descriptive, not comparative. The Other share reflects the registered keyword method's known limitation on titles without genre words (§4.2); the full-corpus developer-tag upgrade is registered future work.**

| Game type | n | Epics | Stories | Skills | Tools | Cost |
|---|---|---|---|---|---|---|
| Other/Unclassified by title | 24 | 15.0 | 55.8 | 3.3 | 4.0 | $0.56 |
| RPG/Fantasy | 6 | 17.5 | 64.0 | 2.4 | 4.4 | $0.61 |
| Horror | 4 | 17.5 | 64.2 | 4.3 | 4.7 | $0.65 |
| Puzzle | 3 | 13.3 | 50.3 | 3.0 | 4.5 | $0.46 |
| Roguelike/Roguelite | 2 | 11.0 | 43.0 | 3.0 | 5.0 | $0.43 |
| Action/Shooter/Fighting | 1 | 20.0 | 74.0 | n/l | n/l | n/l |
| Platformer | 1 | 18.0 | 67.0 | 2.0 | 5.0 | $0.65 |
| Strategy/4X | 1 | 25.0 | 99.0 | 3.0 | 3.0 | $0.94 |